\documentclass[11pt,a4paper]{article}
\usepackage[T1]{fontenc}
\usepackage[utf8]{inputenc}
\usepackage{graphicx}
\usepackage{hyperref}
\usepackage[margin=25mm]{geometry}
\usepackage{microtype}
\usepackage{lmodern}
\usepackage{amsmath}
\usepackage{tabularx}
\usepackage{amssymb}
\hypersetup{
  hidelinks,
  pdftitle={Exceptional points and cusp singularities in relativistic stellar oscillations with causal bulk relaxation},
  pdfauthor={Orlando Urbina-Gonzalez},
  pdfsubject={Relativistic stellar oscillations; causal bulk viscosity; non-Hermitian spectral geometry},
  pdfkeywords={exceptional points; quasinormal modes; Israel-Stewart hydrodynamics; bulk viscosity;
               cusp singularity; hyperboloidal foliation; neutron star mergers}
}

\begin{document}


\title{Exceptional points and cusp singularities in relativistic stellar oscillations with causal bulk relaxation}

\author{Orlando Urbina-Gonzalez\\[2pt]
\small Facultad de F\'isica, Pontificia Universidad Cat\'olica de Chile, Santiago, Chile\\[2pt]
\small \texttt{omurbina@uc.cl}}

\date{}
\maketitle
\begin{center}
\small Published in \emph{Classical and Quantum Gravity} \textbf{43} (2026) 205006.\\
\small \url{https://doi.org/10.1088/1361-6382/aeb05c}\\
\small \copyright\ 2026 The Author. CC BY 4.0: \url{https://creativecommons.org/licenses/by/4.0/}
\end{center}
\vspace{0.6em}

\begin{abstract}
Causal relativistic hydrodynamics promotes dissipative stresses to independent dynamical variables obeying first-order relaxation equations. Coupling such a channel to the second-order wave problem elevates the oscillation spectrum from a quadratic self-adjoint problem to a cubic non-Hermitian one, whose exceptional-point (EP2) curves terminate at an $A_3$ cusp lying in the physically admissible transport domain $(\Lambda>0,\,De>0)$. This geometry turns out to carry a sharp and transferable admissibility criterion: reaching a degeneracy at all requires $(v_{\rm FR}/v_\beta)^2\ge1+\Theta(\gamma)$, where $v_{\rm FR}$ and $v_\beta$ are the frozen and $\beta$-equilibrated sound speeds and $\Theta$ is the minimum of the control ratio over the exceptional-point locus at radiative rate $\gamma$. For the weakly radiating compressive modes to which bulk viscosity couples, $\Theta\simeq8$ and the condition reads $v_{\rm FR}\gtrsim3v_\beta$---a condition on the equation of state alone, independent of the relaxation time, which the internally consistent nucleonic case misses by more than two orders of magnitude, and which no published dense-matter susceptibility examined here comes close to meeting. We regard that criterion, rather than the negative verdict it currently returns, as the transferable result of this work.

Our central result is that this normal form is not merely an effective parametrisation but is exactly realized in a radiating boundary-value problem: a P\"oschl--Teller wave equation coupled to an Israel--Stewart relaxation channel, formulated hyperboloidally so that outgoing (Siegert) conditions become endpoint regularity. We prove a sector-decomposition theorem---the complete quasinormal point spectrum of this open, infinite-dimensional problem, for the aligned profile and away from the discrete resonant couplings, is exactly a tower of cubic normal forms, one per overtone, with each radiative damping rate $\gamma_m$ an output rather than a prescribed constant. We do not claim, and do not need, the absence of essential spectrum for a closed extension to a larger space. Completeness of the point spectrum is established by showing that regularity forces the Frobenius series to truncate, so the polynomial eigenfunctions exhaust it. The EP2 curve, the cusp and its migration law are therefore satisfied exactly by genuine Siegert eigenvalues. Exceptional points are certified by multiprecision bordered-Newton continuation, and the square-root splitting exponent is sampled over eight decades of detuning, the asymptotic fit $0.50000$ being taken over $\epsilon\lesssim10^{-4}$, above which the subleading correction lifts the local slope to $0.5096$.

We then break the exact solvability deliberately: under a controlled one-parameter family of viscosity-profile deformations---parity-preserving and leaving the endpoint indicial structure intact, not a generic perturbation of the operator---the fundamental exceptional point survives a finite window, annihilates with a partner born from a defective inter-sector crossing, and escapes into complexified parameters, converting the sharp overdamping transition into a smooth crossover. The persistence found in this controlled family shows that the exceptional-point skeleton is not solely a consequence of exact solvability, and we delimit the deformation class and the window in which it persists.

Finally, we ask whether dense matter can reach this structure. The control ratio of the two transport parameters is fixed by the equation of state alone, $\Lambda/De=(v_{\rm FR}/v_\beta)^2-1$ with $v_{\rm FR}$ and $v_\beta$ the frozen and $\beta$-equilibrated sound speeds, so that an exceptional point requires $v_{\rm FR}\gtrsim3v_\beta$ for weakly radiating modes, and $(v_{\rm FR}/v_\beta)^2\ge1+\Theta(\gamma)$ in general. Urca-driven $npe\mu$ matter, the one case in which susceptibility and stiffness refer to the same medium, falls short by more than two orders of magnitude, and the Bayesian-allowed space of nucleonic equations of state points the same way. Published hyperonic and colour-superconducting susceptibilities also benchmark one to two orders below threshold, but against the nucleonic fiducial stiffness rather than their own, so they are indicative comparisons and not same-equation-of-state no-go tests. The criterion is independent of the relaxation time and locates the required regime in strongly softened matter.
\end{abstract}

\vspace{6pt}
\noindent\textbf{Keywords:} exceptional points; quasinormal modes; causal bulk viscosity;
Israel--Stewart hydrodynamics; cusp singularity; hyperboloidal foliation; neutron-star oscillations

\section*{Summary for Broad Readership}
When two neutron stars merge, they form a hypermassive remnant that oscillates violently, emitting gravitational waves before potentially collapsing into a black hole. Modern simulations show that the extreme temperatures inside this remnant create "causal bulk friction"---a type of viscosity that takes a finite time to react to the fluid's motion. 

In this work, we show that incorporating this delayed friction changes the mathematical class of the vibration problem, and we exhibit an exactly solvable radiating system---a simplified wave equation with a relaxation channel, with energy genuinely leaking to infinity as it does in a real star---in which the new mathematics can be verified to twenty significant digits analytically, and reproduced to the same twenty by a direct numerical solve of the full operator. That solvable system is a model problem, not a star: it captures the spectral structure exactly, but the step to a fully relativistic stellar calculation remains to be taken. In that model system, instead of simply damping the oscillations, the friction forces the vibrational modes to collide and merge, creating structures known as ``exceptional points'' and ``cusp singularities''---though, as we find below, ordinary neutron-star matter does not appear to reach the regime in which they occur. We then ask whether real neutron-star matter can reach the regime where these structures appear, and derive a sharp criterion for it. Applied to the best current calculations of viscosity from weak nuclear reactions, the answer is negative by a wide margin: ordinary post-merger matter falls short by more than two orders of magnitude. The criterion is nonetheless the useful output, since it can be applied to any candidate form of dense matter and shows that the structures would be reachable only where the matter is anomalously soft, as near a phase transition.

\section{Introduction}

The oscillation spectrum of self-gravitating relativistic fluids is traditionally formulated as a quadratic eigenvalue problem in the mode frequency \cite{Chandrasekhar1964, LindblomDetweiler1983, KokkotasSchmidt1999}. In the absence of dissipation, the associated operator can be rendered self-adjoint under appropriate inner products, leading to a real spectrum for stable configurations and a well-defined modal classification \cite{Kato1966, Chandrasekhar1964}.

Recently, the non-Hermitian nature of gravitational-wave asteroseismology has received intense attention, particularly driven by the spectral instability of black hole quasinormal modes (QNMs) \cite{Berti2009}. Pseudospectrum analyses have revealed that black hole ringdown frequencies are highly sensitive to small environmental or boundary perturbations \cite{Jaramillo2021, Jaramillo2022, Destounis2021, Cheung2022, Boyanov2024}. This structural fragility can lead to the resonant excitation of modes and the generic emergence of exceptional points in the perturbed black hole spectrum \cite{Motohashi2024, Oshita2024, Macedo2025}. The spectral instability in relativistic open systems has emerged as a fundamental feature, as widely discussed in recent literature (see, e.g., the comprehensive review by Shen \textit{et al.} \cite{Shen2025}). However, while non-Hermitian degeneracies in black holes are typically induced by external far-field modifications, this work asks whether comparable degeneracies can arise intrinsically, from internal causal transport rather than from external modifications. We identify the cubic mechanism in the stellar reduction, and we demonstrate the associated exceptional-point and cusp geometry \emph{exactly} in an open radiating scalar model with an Israel--Stewart channel; whether the full relativistic stellar operator realises the same organisation remains open, and is stated as such throughout. We then determine the condition under which dense matter could reach that regime at all.

Causal relativistic hydrodynamics modifies this structure in a fundamental way \cite{Israel1976, IsraelStewart1979}. In Israel--Stewart-type theories, dissipative stresses are promoted to independent dynamical variables obeying first-order relaxation equations \cite{IsraelStewart1979, HiscockLindblom1983}. When coupled to the second-order Einstein--fluid system, this additional temporal order generically elevates the spectral problem from quadratic to cubic in the complex frequency \cite{BhattacharyyaMitraRoy2024, Mitra2025}. The resulting operator pencil is intrinsically non-Hermitian \cite{Kato1966, Heiss2004, Rotter2009}.

While the appearance of non-hydrodynamic relaxation modes in causal theories is well established \cite{HiscockLindblom1985, Muller1967}, the global geometric structure of the resulting stellar oscillation spectrum has not been analysed. In particular, the organisation of eigenvalues in parameter space and the possible emergence of non-Hermitian degeneracies remain unexplored in the context of relativistic stellar oscillations \cite{Berry2004, Heiss2004}.

The dimensionless cubic normal form governing relativistic stellar oscillations with a single bulk relaxation channel possesses a structured set of exceptional points (EP2), forming a critical curve in parameter space \cite{Heiss2004}. Moreover, this curve terminates at a higher-order degeneracy characterised by the simultaneous vanishing of the first and second spectral derivatives, satisfying the standard criteria for an $A_3$ (cusp) singularity in the sense of Arnold’s classification \cite{Arnold1986, ArnoldGuseinZadeVarchenko1985}.

The spectral surface therefore exhibits universal local geometry with square-root splitting near generic exceptional points and cubic-root scaling at the cusp \cite{WangYan2025, BidSchomerus2024}. These non-analytic features arise purely from the structural interplay between second-order gravitational--fluid dynamics and first-order causal relaxation.

The aim of this paper is fivefold. First, we provide a systematic derivation of this spectral geometry at the level of the cubic normal form, including its open-system generalisation. Second---and this is the central result---we prove that the open normal form is \emph{exactly realized} by a genuine radiating boundary-value problem: a hyperboloidally formulated wave--relaxation system with outgoing Siegert boundary conditions whose complete quasinormal point spectrum decomposes into a tower of open cubic polynomials, one per overtone, with radiative damping rates $\gamma_m$ arising as outputs of the eigenvalue problem rather than prescribed parameters (section~\ref{sec:exact_realization}). Third, we certify the resulting exceptional points and the cusp numerically with multiprecision methods, at the level of rigour demanded by non-Hermitian spectral analysis. Fourth, we quantify the structural robustness of this skeleton under a controlled one-parameter family of transport-profile deformations that activates inter-mode coupling, locating explicitly the boundary of the parameter window in which the exceptional-point structure survives on the physical domain. Fifth, we derive a criterion, in terms of the frozen and $\beta$-equilibrated sound speeds alone, for whether a given form of dense matter can reach the exceptional-point regime, and evaluate it against current microphysical calculations (section~\ref{sec:nogo}).

\section{Causal relaxation and cubic spectral structure}
\label{sec:causal_relaxation}

We consider a self-gravitating relativistic fluid described by the Einstein equations coupled to a causal dissipative hydrodynamic theory of Israel--Stewart type \cite{Israel1976, IsraelStewart1979, HiscockLindblom1983}. In such theories, dissipative stresses are promoted to independent dynamical variables obeying first-order relaxation equations \cite{IsraelStewart1979, HiscockLindblom1985}.

For definiteness, the analysis is restricted to a single bulk viscous channel \cite{AlfordMahmoodifarSchwenzer2012, SchmittShternin2018}. The bulk viscous pressure $\Pi$ satisfies a relaxation equation of the form
\begin{equation}
\tau u^\alpha \nabla_\alpha \Pi + \Pi = - \zeta \nabla_\alpha u^\alpha ,
\end{equation}
where $\tau$ is the relaxation time and $\zeta$ the bulk viscosity coefficient \cite{IsraelStewart1979}. The fluid four-velocity is denoted by $u^\alpha$.

The equilibrium configuration is assumed static and spherically symmetric \cite{Chandrasekhar1964}. Linear perturbations are introduced in both the metric and fluid variables. The Einstein equations remain second order in time derivatives \cite{Chandrasekhar1964, LindblomDetweiler1983}, while the fluid equations derived from $\nabla_\mu T^{\mu\nu}=0$ are also second order in time when expressed in terms of the Lagrangian displacement \cite{LindblomDetweiler1983, AnderssonKokkotas1996}.

The relaxation equation, however, is first order in time \cite{IsraelStewart1979, HiscockLindblom1983}. Consequently, after linearisation and Fourier decomposition with time dependence $e^{-i\omega t}$, the coupled system acquires an additional factor of $(1 - i\omega\tau)^{-1}$ in the effective pressure perturbation \cite{HiscockLindblom1985, BhattacharyyaMitraRoy2024}. Because the gravitational--fluid subsystem is second order in time, multiplication by $(1 - i\omega\tau)$ yields a cubic polynomial eigenvalue problem in $\omega$ \cite{Kato1966, TrefethenEmbree2005}.

The resulting spectral equation can be written schematically as
\begin{equation}
\mathcal{P}(\omega) \xi = 0 ,
\end{equation}
where $\mathcal{P}(\omega)$ is a cubic operator polynomial in the complex frequency \cite{Kato1966}.

This elevation of the spectral degree is structural: it follows directly from the temporal differential order of the Einstein--fluid--relaxation system and does not depend on specific microphysical details \cite{BhattacharyyaMitraRoy2024, Mitra2025}. The presence of the relaxation variable renders the operator pencil intrinsically non-Hermitian \cite{Heiss2004, Rotter2009}, defining a universal framework whose spectral consequences resolve into a dimensionless normal form that captures these fundamental properties.

\section{Origin of the cubic spectral structure}

The cubic spectral structure is not a peculiarity of a specific stellar model, gauge choice, or background symmetry. Rather, it follows generically from the differential order of the coupled Einstein--fluid--relaxation system under minimal structural assumptions \cite{IsraelStewart1979, HiscockLindblom1983, Kato1966}.

It is crucial to clarify that this elevation to a cubic pencil is a specific structural hallmark of extended transient thermodynamics (such as Israel--Stewart or M\"uller formulations) \cite{IsraelStewart1979, Muller1967}, which enforce strict causality by promoting dissipative fluxes to independent variables governed by relaxation equations. This differs from alternative modern causal frameworks, such as the Bemfica-Disconzi-Noronha-Kovtun (BDNK) first-order theories \cite{Bemfica2018, Kovtun2019}. In BDNK-type approaches, causality and nonlinear stability are achieved without promoting the dissipative fluxes to independent dynamical variables, but rather through systematic field redefinitions (frame transformations) that absorb the non-equilibrium degrees of freedom into modified inertial parameters \cite{Mitra2025, BhattacharyyaMitraRoy2024}. We are careful not to overstate the contrast: ``first order'' in BDNK refers to the gradient expansion of the constitutive relations, \emph{not} to the temporal differential order of the resulting evolution equations, which do contain time derivatives of the hydrodynamic variables and are in general second order in time, with gapped non-hydrodynamic modes of their own. Indeed Ref.~\cite{BhattacharyyaMitraRoy2024} exhibits frame transformations of BDNK leading to equivalent local formulations with additional non-hydrodynamic degrees of freedom, one of which is third order in time. What is specific to the Israel--Stewart paradigm is therefore not the mere presence of extra temporal order, but the particular coefficient structure that a single scalar relaxation channel imposes on the resulting pencil---the structure analysed here.

\subsection{Temporal differential structure}
\label{sec:temporal_order}

Consider a relativistic self-gravitating fluid whose dynamics is governed by:

\begin{itemize}
\item[(i)] The Einstein equations,
\item[(ii)] Conservation of energy--momentum $\nabla_\mu T^{\mu\nu}=0$,
\item[(iii)] A single scalar relaxation equation of Israel--Stewart type \cite{IsraelStewart1979}.
\end{itemize}

The structural derivation relies on the following assumptions:

\begin{enumerate}
\item The gravitational--fluid subsystem is second order in time derivatives \cite{Chandrasekhar1964, LindblomDetweiler1983}.
\item The relaxation variable satisfies a first-order hyperbolic equation of the form
\[
\tau u^\alpha \nabla_\alpha \Pi + \Pi = \mathcal{S},
\]
with $\tau>0$ and $\mathcal{S}$ depending on fluid variables and their first derivatives \cite{IsraelStewart1979, HiscockLindblom1983}.
\item The background configuration is stationary, so that a Fourier decomposition $e^{-i\omega t}$ is well defined.
\end{enumerate}

Under linearisation about a stationary background, the gravitational--fluid sector produces a quadratic dependence on $\omega$, schematically of the form
\begin{equation}
(-\omega^2 A + i\omega B + C)\,\xi + D\,\Pi = 0,
\label{eq:quadratic_block_struct}
\end{equation}
where $\xi$ collectively denotes gravitational and fluid perturbations \cite{Kato1966}.

The relaxation equation yields
\begin{equation}
(1 - i\omega\tau)\,\Pi = E\,\xi .
\label{eq:relaxation_block_struct}
\end{equation}

Eliminating $\Pi$ from \eqref{eq:quadratic_block_struct} using \eqref{eq:relaxation_block_struct} gives
\begin{equation}
(-\omega^2 A + i\omega B + C)\,\xi
+ \frac{D E}{1 - i\omega\tau}\,\xi = 0.
\end{equation}

Multiplying by $(1 - i\omega\tau)$ produces
\begin{equation}
\left[
(1 - i\omega\tau)(-\omega^2 A + i\omega B + C)
+ D E
\right]\xi = 0,
\end{equation}
which is manifestly cubic in $\omega$.

\medskip
\noindent
\textbf{Proposition.}
\textit{
For any linearised relativistic system in which a second-order hyperbolic gravitational--fluid subsystem is coupled to a single first-order scalar relaxation channel, the reduced spectral problem for normal modes is necessarily of cubic \emph{grade} in the complex frequency $\omega$: the coefficient blocks carry no power of $\omega$ higher than the third, and the cubic leading block is non-vanishing whenever the inertial and relaxation coefficients are non-degenerate. We state it in terms of the grade rather than of the degree of the characteristic determinant because, for a multi-dimensional $\xi$, the latter is not three---an $N$-dimensional problem generically has a determinant of degree $3N$, as section~\ref{sec:functional_reduction} makes explicit.
}
\medskip

\subsection{Physical admissibility and open-system coefficient structure}
\label{sec:open_coefficients}

The cubic polynomial governing the stellar normal modes is not arbitrary; its structure is constrained by strict causal transport and the non-conservative nature of an open gravitational-fluid system. In a realistic astrophysical scenario it is the \emph{perturbation} problem that is open: a static, spherically symmetric background does not radiate, but non-radial perturbations satisfying outgoing conditions lose energy to gravitational radiation, so the baseline hydro-gravitational operator, considered with those boundary conditions, possesses complex quasinormal mode (QNM) roots of the form $\Omega_0 = \pm 1 - i\gamma$, where $\gamma > 0$ is the radiation damping rate. It must be stressed at the outset that in the genuine boundary-value problem $\gamma$ is not a free algebraic parameter: it is an \emph{output} of the nonlinear (Siegert-type) eigenvalue problem, each mode carries its own $\gamma_m$, and radiative and viscous dissipation do not in general factorise. For this reason, in the present section the following one-parameter family of cubics is introduced only as a \emph{definition}---a candidate normal form whose status must be settled at the level of the full boundary-value problem. Precisely this is accomplished in section~\ref{sec:exact_realization}, where the family below is \emph{derived}, exactly and without truncation, as the sector decomposition of the quasinormal spectrum of an open wave--relaxation system with outgoing boundary conditions, in which the $\gamma_m$ emerge as genuine Siegert outputs.

Consider then a quadratic factor with the open-system root pair, $(\Omega - (1 - i\gamma))(\Omega - (-1 - i\gamma)) = \Omega^2 + 2i\gamma\Omega - (1 + \gamma^2)$, coupled to the Israel--Stewart relaxation factor $(1 - iDe\,\Omega)$ and to the bulk viscous attenuation $i\Lambda\Omega$. Expanding the product defines the generalized \textit{Open Cusp Polynomial Form}:
\begin{equation}
P(\Omega; De, \Lambda, \gamma) = -iDe\,\Omega^3 + (1 + 2De\gamma)\Omega^2 + i\left[De(1+\gamma^2) + \Lambda + 2\gamma\right]\Omega - (1+\gamma^2) = 0.
\label{eq:open_polynomial}
\end{equation}

Two remarks delimit the status of Eq.~\eqref{eq:open_polynomial}. First, for a generic star this parametrisation is a single-mode effective model, and its multiplicative structure is an assumption. Second---and this is the decisive point established in section~\ref{sec:exact_realization}---there exists a class of genuinely open systems, formulated with the non-local outgoing (Siegert) boundary conditions that the full gravitational problem requires, whose \emph{exact} quasinormal spectrum is the union of polynomials of precisely the form \eqref{eq:open_polynomial}: one polynomial per overtone $m$, with $\gamma \to \gamma_m$ determined by the boundary-value problem itself. Within that class, every algebraic statement made below about Eq.~\eqref{eq:open_polynomial}---the EP2 curves, the cusp, and its migration law \eqref{eq:deborah_migration}---is a statement about genuine Siegert eigenvalues, exact to all orders. The robustness of these structures outside the exactly solvable class is then a quantitative question, answered in section~\ref{sec:robustness} by certified numerical continuation.

In the asymptotic limit of zero radiation leakage ($\gamma \to 0$), Eq.~\eqref{eq:open_polynomial} smoothly recovers the strictly closed stellar normal form $-iDe\Omega^3 + \Omega^2 + i(De+\Lambda)\Omega - 1 = 0$. The parameter space is bounded by $De > 0$ and $\Lambda > 0$, which fix the signs of the relaxation time and of the bulk viscosity and so ensure positive entropy production. These are necessary conditions, not the whole of physical admissibility: causality of the underlying Israel--Stewart closure is a further, independent restriction, imposed explicitly in section~\ref{sec:transport_scales} through \eqref{eq:IS_causality}. Until then, ``admissible'' refers to the sign conditions alone. Consequently, the presence of $\gamma$ inflates the parameter manifold into a higher-dimensional spectral space, shifting the coordinates of the fundamental degeneracies.

The tension between causality, stability, and sensible initial-value formulations in relativistic dissipative fluids has been the subject of rigorous recent formal developments \cite{Bemfica2018, Kovtun2019, Gavassino2020}. Contemporary analyses, including the Bemfica-Disconzi-Noronha-Kovtun (BDNK) first-order approaches and modern nonlinear stability proofs of extended Israel--Stewart theories \cite{Rischke2024, Cordeiro2025}, reaffirm that the imposition of strict causality fundamentally alters the mathematical class of the differential operators involved. This alteration is what raises the grade of the pencil to three. It does not by itself produce the Riemann-surface topology analysed below---section~\ref{sec:sensitivity} gives an explicit cubic family with two real controls and no degeneracy whatever---which requires in addition the specific coefficient structure that a single scalar relaxation channel imposes.

Crucially, the $A_3$ cusp lies within the physically admissible parameter plane ($\Lambda_* > 0$, $De_* > 0$). This is not automatic: had the organizing center fallen at negative coupling or negative Deborah number, the singularity-theoretic structure would be a formal property of the polynomial family with no bearing on admissible transport. Because it does not, the versal unfolding of the cusp governs a region that thermodynamic admissibility permits, rather than a formal corner of the polynomial family; whether dense matter occupies it is a separate question, settled in the negative in section~\ref{sec:nogo}, and the non-trivial Riemann-surface topology of the spectrum follows from the coefficient structure that a single scalar relaxation channel imposes on the family \eqref{eq:open_polynomial}---not from parameter tuning, but not from the presence of causal relaxation alone either, since section~\ref{sec:sensitivity} gives a cubic family with two real controls and no degeneracy whatever \cite{Arnold1986, Heiss2004}.

\subsection{Formal reduction of the integro-differential operator pencil}
\label{sec:functional_reduction}

To rigorously bridge the gap between the infinite-dimensional stellar interior and the 0-D macroscopic normal form, we must address the spatial heterogeneity of the transport coefficients. In a realistic configuration, the relaxation time $\tau(r)$ and bulk viscosity $\zeta(r)$ are non-uniform functions of the stellar radius. 

Let $\xi^i(x)$ be the Lagrangian displacement field of the fluid. The linearised, purely hydro-gravitational sector of the stellar oscillations (assuming a stable TOV background) can be cast as an abstract eigenvalue problem
\begin{equation}
    (-\omega^2 \mathcal{A} + \mathcal{C}) \xi + \nabla (\delta \Pi) = 0 ,
    \label{eq:pde_system}
\end{equation}
where $\mathcal{A}$ is a positive-definite mass-density operator and $\mathcal{C}$ is the self-adjoint restoring force operator (incorporating pressure gradients and gravitational perturbations) \cite{Chandrasekhar1964, LindblomDetweiler1983, FriedmanStergioulas2013}. The perturbation of the causal bulk viscous pressure, obtained from the Israel--Stewart relaxation equation, is given by the constitutive relation
\begin{equation}
    \delta \Pi = \frac{i \omega \zeta(r)}{1 - i \omega \tau(r)} (\nabla \cdot \xi) .
    \label{eq:delta_pi}
\end{equation}

Inserting \eqref{eq:delta_pi} into \eqref{eq:pde_system} yields a non-linear eigenvalue problem in an infinite-dimensional Hilbert space. Crucially, as the spatial differential operator $\nabla$ does not commute with the rational multiplier $(1-i\omega\tau(r))^{-1}$, simply clearing the denominator across the differential equation is mathematically ill-posed.

To extract the macroscopic effective dynamics, we construct the weak formulation. We define the standard inner product $\langle \eta, \xi \rangle = \int_\Sigma \rho \eta_i^* \xi^i dV$ over the stellar volume $\Sigma$. Projecting the physical spatial differential equation \eqref{eq:pde_system} onto a test vector field $\eta^i$ and integrating the viscous pressure gradient term by parts yields a volume and a surface contribution. The physical boundary of the problem, corresponding to the outer stellar crust $\partial\Sigma$, imposes Neumann-like boundary conditions where the density and the macroscopic bulk viscous stress $\delta\Pi$ vanish, so that the surface flux integral is eliminated. We flag this as a hypothesis rather than a property shared by every relativistic star: the physical condition at a stellar surface is normally imposed on the Lagrangian perturbation of the pressure or traction, which does not universally force $\delta\Pi=0$, and for self-bound configurations the surface density need not vanish at all. The reduction below is valid for the class of boundary conditions that annihilate the surface form, and we restrict to that class. Consequently, the convolution reduces purely to the bulk divergence, resolving into:
\begin{equation}
-\omega^2 \langle \eta, \mathcal{A} \xi \rangle + \langle \eta, \mathcal{C} \xi \rangle - \left\langle \nabla \cdot \eta, \frac{i \omega \zeta(r)}{1 - i \omega \tau(r)} \nabla \cdot \xi \right\rangle = 0 .
\label{eq:weak_form}
\end{equation}

We now perform a single-mode Galerkin truncation \cite{Kato1966}. Let $\xi_0$ be the fundamental eigenvector of the non-dissipative system, satisfying $\mathcal{C} \xi_0 = \omega_0^2 \mathcal{A} \xi_0$. Setting $\xi = \xi_0$ and the test function $\eta = \xi_0$, the non-dissipative terms reduce to the quadratic kinetic energy $\langle \xi_0, \mathcal{A} \xi_0 \rangle (\omega_0^2 - \omega^2)$. 

For the dissipative integral, the spatial convolution can be written in terms of an effective frequency-dependent relaxation pair $(\tilde{\tau}(\omega), \tilde{\zeta}(\omega))$. This step is by itself a tautology and we do not present it as more: at each $\omega$ the integral is a single complex number, matched by infinitely many such pairs. The step carrying physical content is the next one, and it is an approximation rather than an identity:
\begin{equation}
    \int_\Sigma \frac{i \omega \zeta(r)}{1 - i \omega \tau(r)} |\nabla \cdot \xi_0|^2 dV \equiv \frac{i \omega \tilde{\zeta}(\omega)}{1 - i \omega \tilde{\tau}(\omega)} \int_\Sigma |\nabla \cdot \xi_0|^2 dV .
\end{equation}
In the weak-gradient regime of the transport coefficients across the support of the fundamental mode, these frequency-dependent functions are \emph{replaced} by constant macroscopic parameters $\bar{\tau}$ and $\bar{\zeta}$. We do not derive an error bound for that replacement in terms of $\nabla\tau$, $\nabla\zeta$ and the modal support, and accordingly we call the result a constant-coefficient single-mode approximation rather than a controlled reduction with a quantified small parameter. What the argument establishes rigorously is the grade of the projected pencil; the numerical value of the effective coefficients is a modelling input.

Substituting this parametrisation into \eqref{eq:weak_form}, dividing by the unperturbed kinetic energy $\langle \xi_0, \mathcal{A} \xi_0 \rangle$, and clearing the macroscopic scalar denominator $(1 - i \omega \bar{\tau})$, we obtain exactly the 0-D cubic operator pencil:
\begin{equation}
    (1 - i \omega \bar{\tau})(\omega_0^2 - \omega^2) - i \omega \frac{\bar{\zeta} \int |\nabla \cdot \xi_0|^2 dV}{\langle \xi_0, \mathcal{A} \xi_0 \rangle} = 0 .
\label{eq:pencil_reduced}
\end{equation}

It is important to note that the parameter $\Lambda$ encapsulates the spatial momentum dependence of the hydrodynamic modes. In the short-wavelength (WKB) limit, the spatial integral scales as $\int_\Sigma |\nabla \cdot \xi_0|^2 dV \propto k_{eff}^2$. This demonstrates that our dimensionless parameter formally scales as $\Lambda \propto k_{eff}^2 \zeta$, recovering the standard local hydrodynamic dispersion relation where viscous attenuation depends on the spatial momentum, discretized in a bounded star by its multipolar and radial numbers. 

We explicitly clarify that while the WKB geometric optics approximation strictly requires the perturbation wavelength to be much smaller than the pressure scale height ($\lambda \ll R$), its presentation here for the macroscopic composition $g$-mode serves an illustrative and conceptual role to map spatial diffraction into a 0-D representation. A completely rigorous treatment of the global $g$-mode viscous coupling inherently requires computing non-local volumetric projection coefficients rather than a local wave-number scaling. Nevertheless, the grade of the projected pencil remains cubic under such projections. We do not extend that invariance to the Riemann-surface topology: section~\ref{sec:sensitivity} exhibits a cubic family with two real controls whose discriminant never vanishes, so the EP and cusp geometry does not follow from the grade alone but from the coefficient structure, which a multi-mode projection may alter.

By defining the dimensionless frequency $\Omega = \omega / \omega_0$, the Deborah number $De = \omega_0 \bar{\tau}$, and the viscous coupling parameter
\begin{equation}
    \Lambda = \frac{\bar{\zeta}}{\omega_0 \langle \xi_0, \mathcal{A} \xi_0 \rangle} \int_\Sigma |\nabla \cdot \xi_0|^2 dV ,
\end{equation}
the infinite-dimensional PDE reduces, under the single-mode Galerkin projection and the constant-coefficient approximation declared above, to our dimensionless cubic normal form \eqref{eq:cubic}. 

Because the cubic order originates fundamentally from the product of a second-order hyperbolic operator (the Einstein--fluid sector) with a first-order relaxation operator, the \emph{grade} of the matrix polynomial---the highest power of $\omega$ carried by the coefficient blocks---remains at most three under a non-degenerate finite-dimensional Galerkin projection, with a non-vanishing cubic leading block whenever the projected inertial and relaxation coefficients are non-degenerate. We state it in terms of the grade rather than the degree of the characteristic determinant, because the latter is \emph{not} invariant: an $N$-dimensional projection of a cubic pencil has a determinant of degree $3N$ generically, as the diagonal example $\mathrm{diag}(\omega^3+1,\,\omega^3+2)$ already shows. This Galerkin derivation demonstrates that the 0-D algebraic representation is not an ad-hoc construct, but a formal topological skeleton of the physical mode dynamics. Whether off-diagonal spatial heterogeneities $\nabla_r \tau(r)$ in a full multi-mode expansion fragment the exact $A_3$ cusp singularity is not settled by this argument; it is answered quantitatively, for a controlled deformation of exactly this type, in section~\ref{sec:robustness}, where the cusp is found to persist over a finite window. What the degree count above does establish is narrower, and we state it as such: the grade of the projected pencil remains cubic under non-degenerate projections. Whether that cubic pencil is \emph{defective} is a separate question---a cubic pencil is diagonalisable away from its discriminant locus---and it is settled not by the degree count but by the coefficient structure analysed in sections~\ref{sec:cusp}--\ref{sec:EP}.

\section{Dimensionless normal form}
To rigorously define the baseline topological skeleton before introducing the radiative open-system deformation, we first analyze the strictly conservative hydrodynamic limit ($\gamma = 0$). In this regime, the cubic spectral structure derived in the previous section admits an expression in a dimensionless normal form that captures its universal features independently of detailed stellar models \cite{Kato1966, TrefethenEmbree2005}.

After suitable rescaling of the frequency and background quantities, the spectral problem reduces to a cubic polynomial equation of the form
\begin{equation}
- i De \,\Omega^3 + \Omega^2 + i(De + \Lambda)\Omega - 1 = 0 ,
\label{eq:cubic}
\end{equation}
where $\Omega$ is a dimensionless complex frequency defined as
\begin{equation}
\Omega = \omega / \omega_0 ,
\end{equation}
with $\omega_0$ denoting the characteristic oscillation frequency of the corresponding non-dissipative configuration \cite{Chandrasekhar1964, LindblomDetweiler1983}.

For later use, this defines explicitly the polynomial
\begin{equation}
P(\Omega; De, \Lambda) = - i De \,\Omega^3 + \Omega^2 + i(De + \Lambda)\Omega - 1 .
\label{eq:polynomial}
\end{equation}

The two dimensionless parameters are defined as follows. The first,
\begin{equation}
De = \omega_0 \tau,
\end{equation}
is the Deborah number, measuring the ratio between the relaxation time $\tau$ and the characteristic dynamical timescale $\omega_0^{-1}$ \cite{BirdArmstrongHassager1987}. The second,
\begin{equation}
\Lambda = \frac{\zeta}{\rho \mathcal{L}^2 \omega_0},
\label{eq:Lambda_def}
\end{equation}
is a dimensionless viscous coupling parameter, equivalent to the volumetric functional $\Lambda = \bar{\zeta}\,[\int_\Sigma |\nabla\cdot\xi_0|^2 dV]/[\omega_0\langle\xi_0,\mathcal{A}\xi_0\rangle]$ derived in section~\ref{sec:functional_reduction}. Here, $\rho$ is the effective inertia per unit volume of the mode---the characteristic density appearing in the kinetic-energy normalisation $\langle\xi_0,\mathcal{A}\xi_0\rangle$, evaluated in section~\ref{sec:transport_scales} as the mean density---and $\mathcal{L} \equiv k_{eff}^{-1}$ represents the effective macroscopic length scale of the fundamental mode, explicitly linking the macroscopic length scale to the momentum of the hydrodynamic perturbation \cite{AlfordMahmoodifarSchwenzer2012, SchmittShternin2018}.

Any linearised Einstein--fluid system supplemented by a single scalar first-order relaxation channel admits a corresponding cubic normal-form reduction under the single-mode Galerkin projection, up to redefinitions of $De$ and $\Lambda$ \cite{IsraelStewart1979, HiscockLindblom1983}. The cubic structure therefore represents a generic effective normal form for relativistic stellar oscillations.

The polynomial is intrinsically non-Hermitian due to the imaginary coefficients proportional to $De$ \cite{Heiss2004, Rotter2009}. In the limit $De \to 0$, the cubic term disappears and the problem reduces to a quadratic eigenvalue equation corresponding to the Navier--Stokes regime ($\Omega^2 + i\Lambda\Omega - 1 = 0$), defining a non-self-adjoint operator with a strictly dissipative imaginary part.

The spectral surface defined by equation \eqref{eq:polynomial},
\begin{equation}
\mathcal{S} = 
\left\{
(\Omega, De, \Lambda) \in \mathbb{C} \times \mathbb{R}^2 
    \; \big| \;
P(\Omega; De, \Lambda) = 0
\right\},
\end{equation}
is therefore a three-sheeted complex surface over the parameter plane $(De,\Lambda)$, except at parameter values where distinct sheets coalesce \cite{Arnold1986, Heiss2004}. The analysis of these degeneracies characterizes the global topology of the system.

\section{Global discriminant structure of the cubic family}

The spectral topology of a polynomial eigenvalue problem is governed globally by its discriminant \cite{Arnold1986, Kato1966}. For a cubic polynomial
\[
a \Omega^3 + b \Omega^2 + c \Omega + d = 0,
\]
the discriminant is given by
\[
\Delta = 
18abcd - 4b^3 d + b^2 c^2 - 4ac^3 - 27a^2 d^2.
\]

For the dimensionless normal form
\[
- i De \,\Omega^3 + \Omega^2 + i(De + \Lambda)\Omega - 1 = 0,
\]
we identify
$a=-iDe$, $b=1$, $c=i(De+\Lambda)$, $d=-1$.

Substituting into the general expression yields a real-valued function
\[
\Delta(De,\Lambda),
\]
which defines a curve in the parameter plane via the condition
\[
\Delta(De,\Lambda) = 0.
\]

\subsection{Exceptional points and the discriminant locus}
\label{sec:discriminant_EP}

For polynomial families, the condition $\Delta = 0$ is equivalent to the existence of a multiple root \cite{Kato1966}. Therefore, the exceptional-point (EP2) curve derived previously coincides precisely with the discriminant locus of the cubic family. This establishes the EP curve as the global algebraic degeneracy set of the two-parameter family, rather than a sectorial artifact. It is worth situating this structure within relativistic hydrodynamics at large: collisions of hydrodynamic and non-hydrodynamic modes at critical points of complexified spectral curves are known to set the convergence radii of the gradient expansion in Israel--Stewart-type and holographic theories \cite{Grozdanov2019PRL, Grozdanov2019JHEP}. The exceptional-point curve derived here is the real-parameter incarnation of that mechanism for a self-gravitating configuration, realised in section~\ref{sec:exact_realization} by genuine Siegert eigenvalues and organised globally by the $A_3$ center.

\subsection{Higher-order degeneracy and cusp condition}
\label{sec:cusp}

A higher-order degeneracy occurs when, in addition to $\Delta = 0$, the gradient of the discriminant also vanishes, $\nabla \Delta = 0$.

This condition identifies a singular point of the discriminant curve in parameter space. For cubic families, such singular points correspond to $A_3$ cusp singularities in the sense of Arnold’s classification \cite{Arnold1986}. A word on convention, since two are in circulation. Here $A_3$ is the cusp catastrophe of the potential $V(x)=x^4+ux^2+vx$, whose equilibrium equation $V_x=0$ is a cubic---the family we analyse. In the strict ADE labelling of function germs, where $A_k\sim x^{k+1}$, the germ $P\sim\delta^3$ would itself be called $A_2$. We follow the catastrophe-theoretic convention throughout: the object carrying the $A_3$ label is the potential, and $P=0$ is its associated equilibrium equation.

Direct evaluation confirms that the triple degeneracy previously identified satisfies
\[
\Delta(De_*,\Lambda_*) = 0,
\qquad
\nabla \Delta(De_*,\Lambda_*) = 0,
\]
while the Hessian of $\Delta$ there has rank one, its kernel lying along the tangent to the discriminant curve. This degeneracy is the signature of the cusp rather than a defect of it: we state the contrast carefully, since its naive form is false. A non-degenerate Hessian does \emph{not} by itself make the zero set an isolated point---that additionally requires a definite signature, and an indefinite non-degenerate Hessian gives two branches crossing transversally, as $x^2-y^2$ does. What distinguishes the present case is the \emph{degenerate} Hessian, of rank one, with kernel along the common tangent of the two branches: they meet tangentially rather than transversally. Explicitly, at the closed-limit cusp $H=\begin{pmatrix}96&24\\24&6\end{pmatrix}$, with kernel $(1,-4)$, matching the tangent direction $(De''(\sqrt3),\Lambda''(\sqrt3))\propto(1,-4)$ obtained from \eqref{eq:EP_param}. 

The cusp is therefore a global singularity of the discriminant locus and not merely a local feature of a restricted spectral sector.

\subsection{Analytic Location of the Open-System Organizing Center}
\label{sec:organizing_center}

To map the global non-Hermitian topology under Arnold's paradigm, the higher-order singularity where three spectral sheets coalesce into a single root must be located. A strict $A_3$ cusp degeneracy requires the simultaneous vanishing of the polynomial and its first two spectral derivatives: $P = P_\Omega = P_{\Omega\Omega} = 0$. 

Differentiating the open polynomial Eq.~\eqref{eq:open_polynomial} twice with respect to the complex frequency $\Omega$ isolates the high-frequency hyperbolic behavior:
\begin{equation}
P_{\Omega\Omega} = -6iDe\,\Omega + 2(1 + 2De\gamma) = 0,
\end{equation}
which enforces the critical degenerate frequency to shift along the imaginary axis as:
\begin{equation}
\Omega_* = -i\frac{1 + 2De\gamma}{3De}.
\label{eq:critical_omega_open}
\end{equation}
Substituting Eq.~\eqref{eq:critical_omega_open} into the first derivative condition $P_\Omega = 0$ allows us to isolate the exact open-system critical viscous coupling:
\begin{equation}
\Lambda_* = \frac{(1+2De\gamma)^2}{3De} - De(1+\gamma^2) - 2\gamma.
\label{eq:critical_lambda_open}
\end{equation}

Finally, substituting both $\Omega_*$ and $\Lambda_*$ back into the master characteristic equation $P(\Omega_*) = 0$ yields the fundamental algebraic constraint that determines the migration of the critical Deborah number $De_*$ driven by the radiation field:
\begin{equation}
(1 + 2De_*\gamma)^3 = 27De_*^2(1 + \gamma^2).
\label{eq:deborah_migration}
\end{equation}
Eq.~\eqref{eq:deborah_migration} proves that the strict $A_3$ organizing center is not destroyed by open-system dissipation; rather, it migrates along a continuous algebraic trajectory, and it does so \emph{without leaving the admissible quadrant}: on the primary branch $w=\gamma De_*\to1^-$ and $\Lambda_*\to0^+$ as $\gamma\to\infty$, so $De_*>0$ and $\Lambda_*>0$ for every radiative rate. For a strictly conservative star without radiation leakage ($\gamma = 0$), the algebraic relation explicitly collapses to $1 = 27De_*^2$, which immediately yields $De_* = 1/(3\sqrt{3})$, perfectly recovering the structural skeleton of our closed-system limit.

\medskip
\noindent\textbf{Lemma (Uniqueness of the admissible open branch).}
\textit{Restrict to the physically admissible domain $0 < \gamma De_* < 1$, and define $w \equiv \gamma De_* \in (0,1)$. Equation (23) can be rewritten as $(1+2w)^3/w^2 = 27(1+\gamma^2)/\gamma^2$. The derivative of the left-hand side is $2(w-1)(2w+1)^2 / w^3$, which is strictly negative on $w \in (0,1)$, meaning the function is strictly decreasing. Since the right-hand side is strictly greater than $27$ for any $\gamma > 0$, the equation admits exactly one unique solution in the admissible domain $0 < \gamma De_* < 1$. This proves that the primary branch continuously connected to the closed-system limit is mathematically unique.}
\medskip

\noindent
\textbf{Proposition (Persistence of the $A_3$ Organizing Center).}
\textit{Along the one-parameter family of radiative deformations parametrised by $\gamma$, restricted to $De_*\gamma < 1$, the critical stratum $(De_*,\Lambda_*,\Omega_*)$ persists, the local unfolding remains diffeomorphic to Arnold's $A_3$ normal form, and the codimension-two cusp structure is preserved.}

\noindent \textit{Proof.} The local persistence rests on three structural pillars established in this work:
(i) \textit{Hyperbolic preservation:} The parameter $\gamma$ only modifies the lower-order coefficients of the characteristic polynomial (equation \ref{eq:open_polynomial}) and cannot annihilate the leading hyperbolic term $(-iDe\Omega^3)$ for any $De>0$, preserving the dimension of the operator pencil.
(ii) \textit{Stratum existence:} The continuous migration and existence of the primary critical stratum $(De_*(\gamma), \Lambda_*(\gamma), \Omega_*(\gamma))$---continuously connected to the closed-system limit---is demonstrated constructively. \\
(iii) \textit{Transversality:} As formally proven in Appendix A, the transformation to Arnold's canonical control space maintains a strictly negative Jacobian determinant ($\det(J) < 0$, equation~\eqref{eq:final_real_determinant}). By satisfying the transversality condition of the Thom-Arnold versal unfolding theorem \cite{Arnold1986}, this ensures a regular local diffeomorphism, formally confirming that the local unfolding preserves the exact $A_3$ structure provided the open-system deformation satisfies $De_*\gamma < 1$. $\blacksquare$

\medskip
\noindent\textbf{Remark.} The Lemma and the Proposition above are statements about the polynomial family \eqref{eq:open_polynomial}. Their physical content is secured by section~\ref{sec:exact_realization}: in the exactly solvable open system constructed there, the migration law \eqref{eq:deborah_migration} is satisfied by the genuine Siegert spectrum with residual zero to forty decimal digits, and the parameter $\gamma$ is swept continuously by the depth of the background potential and discretely by the overtone index---both physically meaningful deformations of the boundary-value problem, not abstract coefficient changes.

\subsection{Topological interpretation}

The discriminant curve partitions the admissible parameter plane into regions with distinct spectral topology. Since the coefficients of the normal form \eqref{eq:open_polynomial} are complex, the classical real-coefficient sign rule does not apply directly to $\Omega$; the correct statement is obtained in the variable $y$ defined by $\Omega=-iy$. It is worth emphasising that this is not a device confined to the closed limit: for \emph{every} real $\gamma$ the substitution yields
\begin{equation}
P(-iy) \;=\; De\,y^3 - (1+2De\,\gamma)\,y^2 + \big[De(1+\gamma^2)+\Lambda+2\gamma\big]\,y - (1+\gamma^2),
\label{eq:open_real_form}
\end{equation}
whose coefficients are manifestly real, so the conjugation symmetry of the spectrum, the reality of the discriminant and the real-coefficient analysis below extend verbatim to the open normal form. In the closed limit this reduces to the polynomial whose discriminant is given above. Note that the discriminant of the cubic in $y$ is $\Delta_y=-\Delta$, so the familiar real-coefficient sign rule appears here with the opposite sign convention. In terms of the physical modes:

\begin{itemize}
\item Regions where $\Delta<0$: all three roots are purely imaginary in $\Omega$ (three distinct real $y$), i.e.\ three non-propagating, overdamped modes.
\item Regions where $\Delta>0$: one purely imaginary root together with a propagating pair symmetric about the imaginary axis (one real $y$ and a complex-conjugate pair), i.e.\ one overdamped mode and one damped oscillatory doublet.
\end{itemize}

Note that no root is ever real in $\Omega$: in the dissipative regime every mode is damped, and the transition encoded by $\Delta$ is between propagating and non-propagating behaviour, not between real and complex frequencies.

Crossing the discriminant curve changes the algebraic multiplicity structure and reorganises the branching of the associated Riemann surface \cite{Heiss2004, Rotter2009}.

The cusp marks the point at which the EP curve changes curvature and the square-root branching characteristic of generic exceptional points transitions to cubic-root behaviour. This global discriminant structure therefore encodes the full non-Hermitian topology of the cubic spectral family.

\section{Exceptional points of the cubic spectrum}
\label{sec:EP}

Having identified the discriminant $\Delta(De,\Lambda)$ as the global algebraic invariant governing spectral degeneracies, an explicit parametrisation of its physically relevant branch follows directly.

For polynomial eigenvalue problems, the condition $\Delta=0$ is equivalent to the simultaneous vanishing of the polynomial and its first derivative \cite{Kato1966}. Exceptional points therefore satisfy
\begin{equation}
P(\Omega; De, \Lambda) = 0 ,
\label{eq:EP_condition1}
\end{equation}
\begin{equation}
\partial_\Omega P(\Omega; De, \Lambda) = 0 .
\label{eq:EP_condition2}
\end{equation}

\subsection{Physically relevant sector}

The analysis focuses on the generalized sector in which the degenerate frequency is purely imaginary,
\begin{equation}
\Omega = - i y ,
\qquad
y > 0 ,
\label{eq:imaginary_sector}
\end{equation}
corresponding to overdamped relaxation modes \cite{HiscockLindblom1983}. This is not in fact a restriction, and saying so costs one line. By \eqref{eq:open_real_form} the polynomial in $y$ has real coefficients for every real $(De,\Lambda,\gamma)$, and a real cubic cannot have a non-real double root: the conjugate would be a second double root, forcing degree four. Every degeneracy of the family therefore lies on $\Omega=-iy$ with $y$ real. Moreover $y<0$ would give $\Lambda=[(y-\gamma)^2+1]^2/[2y^2(y-\gamma)]<0$, outside the admissible quadrant. The parametrisation below is consequently the \emph{complete} exceptional-point locus of the family on the admissible domain, not a sector of it---consistently with section~\ref{sec:discriminant_EP}, where the same conclusion is reached from the discriminant.

Substituting \eqref{eq:imaginary_sector} into \eqref{eq:polynomial} yields
\begin{equation}
P(-iy; De, \Lambda) = De\, y^3 - y^2 + (De + \Lambda) y - 1 ,
\label{eq:P_real}
\end{equation}
which is real-valued.

Similarly, the derivative condition becomes
\begin{equation}
\partial_\Omega P(-iy; De, \Lambda) = i \Big[ 3 De\, y^2 - 2 y + (De + \Lambda) \Big] .
\label{eq:dP_real}
\end{equation}

The exceptional-point condition therefore reduces to the real non-linear algebraic system:
\begin{equation}
\begin{aligned}
De\, y^3 - y^2 + (De + \Lambda) y - 1 &= 0, \\
3 De\, y^2 - 2 y + (De + \Lambda) &= 0.
\end{aligned}
\label{eq:EP_system}
\end{equation}

Solving \eqref{eq:EP_system} yields the explicit parametrisation
\begin{equation}
De(y) = \frac{y^2 - 1}{2 y^3},
\qquad
\Lambda(y) = \frac{(y^2 + 1)^2}{2 y^3}.
\label{eq:EP_param}
\end{equation}

As $y$ varies over the overdamped domain $y > 1$, equations \eqref{eq:EP_param} trace the locus of second-order exceptional points (EP2) of the cubic spectrum back directly into the positive parameter sector \cite{Heiss2012}. This shows that the branch of exceptional points lying strictly on the purely imaginary frequency axis is firmly located in the physically admissible half-plane ($\Lambda > 0$). The stable EP2 curves therefore lie in the thermodynamically admissible quadrant $\Lambda>0$, $De>0$. Whether dense matter occupies the relevant part of that quadrant is a separate question, settled in section~\ref{sec:nogo}.

\begin{figure}[htbp]
\centering
\includegraphics[width=0.75\linewidth]{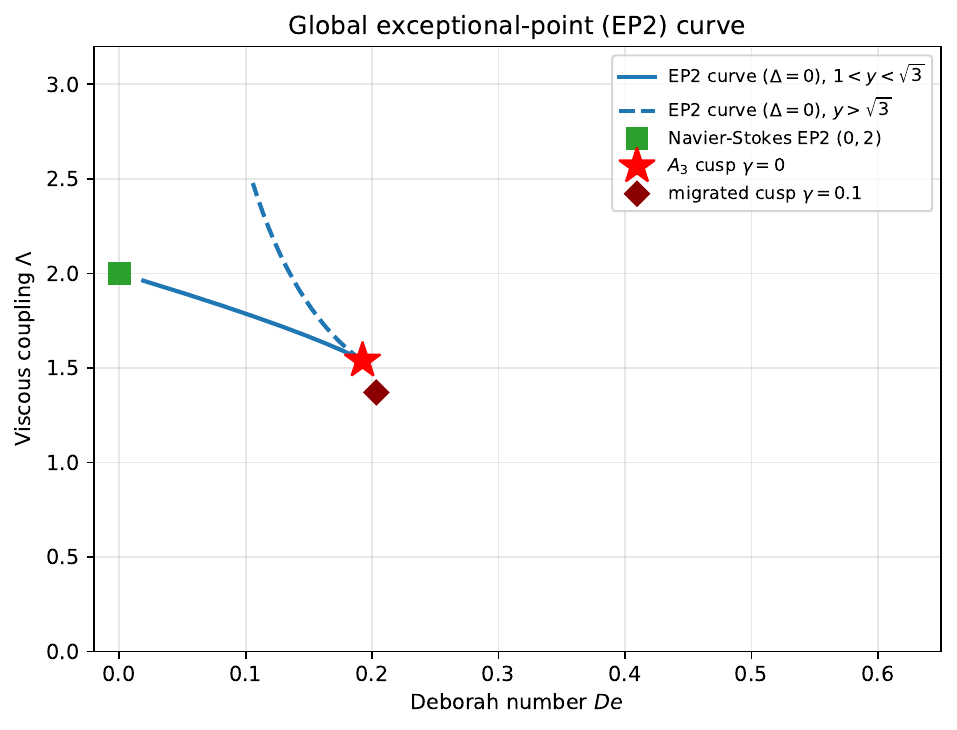}
\caption{Global exceptional-point (EP2) curve---shown here in a truncated window, since the upper branch continues to $De\to0$ with $\Lambda\to\infty$ and cannot be displayed in full---in the generalized sector $\Omega=-iy$, obtained from the exact parametrisation of the discriminant locus $\Delta=0$ (equations (30)). The locus is parametrised by $y=|\Omega|$ through \eqref{eq:EP_param} and has two branches, since $De(y)$ attains its maximum at $y=\sqrt3$, which is the cusp. On the lower branch ($1<y<\sqrt3$, solid) the viscous coupling required for root coalescence decreases monotonically from the Navier--Stokes value $\Lambda=2$ toward the cusp value $\Lambda_*=8/(3\sqrt{3})\approx1.5396$ as $De$ increases; the upper branch ($y>\sqrt3$, dashed) returns from the cusp toward $De\to0$ with $\Lambda\to\infty$. The two branches meet at the cusp with a common tangent, in the direction $(1,-4)$ computed in section~\ref{sec:cusp}. The curve meets the Navier-Stokes parabolic boundary at $(De,\Lambda)=(0,2)$, which represents a regular secondary exceptional point (EP2) where the two propagating fluid modes coalesce, rather than the true cusp. The cusp marker corresponds to the closed limit $\gamma=0$, $(De_*,\Lambda_*)=(1/(3\sqrt{3}),\,8/(3\sqrt{3}))\approx(0.1925,\,1.5396)$; for finite radiative damping it migrates according to Eq.~\eqref{eq:deborah_migration}, e.g.\ to $(0.2033,\,1.3704)$ at $\gamma=0.1$.}
\label{fig:ep_curve}
\end{figure}

Physically, the intersection of this locus with the parabolic boundary $(De=0)$ at $\Lambda=2$
marks the regular Navier-Stokes exceptional point where the non-causal propagating fluid modes coalesce. In our open-system topology, this boundary point serves as the physical gateway from which the complex branch of secondary exceptional degeneracies unfolds, driven continuously by the non-zero relaxation time into the physical transport domain. The resulting curve constitutes the locus of second-order exceptional points (EP2) of the cubic spectrum, as shown in figure~\ref{fig:ep_curve}.

\subsection{Reduction to normal form}

The local structure of the spectral surface near the triple degeneracy is governed by an asymmetric deployment. Let
\begin{equation}
\Omega = \Omega_* + \delta ,
\qquad
De = De_* + \mu ,
\qquad
\Lambda = \Lambda_* + \nu .
\end{equation}

Since $P = P_\Omega = P_{\Omega\Omega} = 0$ at the critical point, the Taylor expansion of $P$ about $(\Omega_*, De_*, \Lambda_*)$ begins at cubic order in $\delta$ in the unperturbed direction, with linear coupling to the control parameters $(\mu,\nu)$ at \emph{two} orders in $\delta$. Because $P$ is exactly linear in $De$ and in $\Lambda$, the expansion terminates and can be written without remainder in those variables:
\begin{equation}
P(\Omega;De,\Lambda)
=
\Big(\tfrac{1}{6} P_{\Omega\Omega\Omega} + \tfrac16 P_{\Omega\Omega\Omega De}\,\mu\Big)\,\delta^3
+ \tfrac12 P_{\Omega\Omega De}\,\mu\,\delta^2
+ \big(P_{\Omega De}\,\mu + P_{\Omega\Lambda}\,\nu\big)\,\delta
+ \big(P_{De}\,\mu + P_{\Lambda}\,\nu\big) ,
\label{eq:local_expansion}
\end{equation}
an exact identity with no remainder whatever. Three facts make it exact. $P$ is a cubic polynomial in $\Omega$, so there is no $\delta^4$ term; $P$ is affine in $(De,\Lambda)$, so no term is quadratic in the controls; and $P_{\Omega\Omega\Lambda}=0$ identically, so there is no $\delta^2\nu$ term. Note that the cubic coefficient itself carries a $\mu$-dependence, since the leading coefficient of $P$ is $-iDe$: at the closed cusp $\tfrac16P_{\Omega\Omega\Omega De}=-i$, so the $\delta^3$ coefficient is $-i(\tfrac{\sqrt3}{9}+\mu)$ and not $-i\tfrac{\sqrt3}{9}$. This term is of higher order in the unfolding and does not affect the $A_3$ classification---it merely rescales $\delta^3$ by a factor $1+\mathcal{O}(\mu)$, non-vanishing near $\mu=0$---but the identity is not exact without it. Dividing through by that coefficient and removing the $\delta^2$ term by a Tschirnhaus translation of the spectral coordinate depresses the cubic \emph{exactly}, so the reduction to the normal form below involves no truncation whatever. The retained mixed linear terms are essential and must not be dropped: they are precisely what produces the coefficient $u$ of the normal form \eqref{eq:normal_form} below. They do not vanish---$P_{\Omega\Lambda}=i$ identically---and at the closed-limit cusp $(\Omega_*,De_*,\Lambda_*)=(-i\sqrt3,\,1/(3\sqrt3),\,8/(3\sqrt3))$ one has, explicitly,
\begin{equation}
\tfrac16 P_{\Omega\Omega\Omega}=-\tfrac{i\sqrt3}{9},\quad \tfrac16 P_{\Omega\Omega\Omega De}=-i,\qquad
P_{\Omega De}=10i,\quad P_{\Omega\Lambda}=i,\qquad
P_{De}=4\sqrt3,\quad P_{\Lambda}=\sqrt3,
\label{eq:cusp_derivatives}
\end{equation}
so that, to the order relevant for the unfolding,
\begin{equation}
P \;=\; -\tfrac{i\sqrt3}{9}\,\delta^3 \;+\; i\,(10\mu+\nu)\,\delta \;+\; \sqrt3\,(4\mu+\nu) \;+\;\cdots
\label{eq:cusp_unfolding_explicit}
\end{equation}
The $\delta^2\mu$ term, retained in the remainder above, is removable by a smooth translation of the spectral coordinate and does not affect the classification. Appendix~A carries out the same expansion at finite $\gamma$ and keeps the mixed terms in the form $(P^*_{\Omega De}\mu+P^*_{\Omega\Lambda}\nu)\,i\tilde\delta$; \eqref{eq:local_expansion} is its closed-limit specialisation.

Because $P_{\Omega\Omega\Omega} \ne 0$ at the critical point, the gradient $(P_{De}, P_{\Lambda})$ is non-vanishing, and the pair $(P_{\Omega De},P_{\Omega\Lambda})$ is independent of it---at the closed cusp the two rows $(4\sqrt3,\sqrt3)$ and $(10i,i)$ have determinant $-6\sqrt3\,i\neq0$---the unfolding depends on two independent control parameters \cite{Arnold1986}. After an invertible linear redefinition of $(\mu,\nu)$, the local equation reduces to the standard cubic normal form
\begin{equation}
\delta^3 + u \delta + v = 0 ,
\label{eq:normal_form}
\end{equation}
where $(u,v)$ are linear combinations of $(\mu,\nu)$.

\begin{figure}[htbp]
\centering
\includegraphics[width=0.75\linewidth]{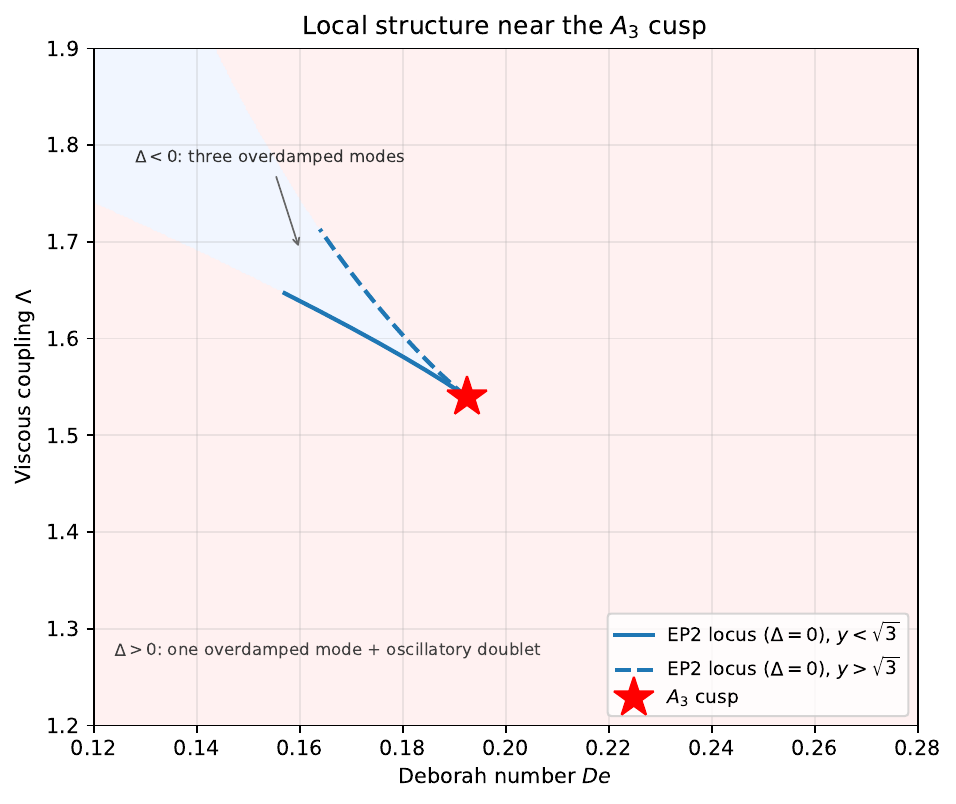}
\caption{Local geometric structure of the discriminant locus near the higher-order degeneracy. Close-up of the parameter plane $(De,\Lambda)$, shaded by the sign of the discriminant. The region $\Delta<0$, in which all three modes are overdamped and non-propagating, is the narrow wedge between the two branches of the locus rather than a half-plane, in the window shown. The true $A_3$ cusp organizing center natively resides within the physical viscosity domain, acting as the organizing center of the locus. The versal unfolding is a \emph{local} statement about the neighbourhood of the cusp; the arrangement of the locus across the quadrant is known here because the cubic family is solvable in closed form, not as a consequence of the local unfolding alone.}
\label{fig:cusp_zoom}
\end{figure}

The local arrangement of the two branches and of the discriminant regions is shown in figure~\ref{fig:cusp_zoom}. The unfolding therefore has codimension two, as required for an $A_3$ singularity in Arnold’s classification \cite{Arnold1986}. The discriminant of \eqref{eq:normal_form},
\begin{equation}
\Delta = -4 u^3 - 27 v^2 ,
\end{equation}
vanishes along a cusp curve in parameter space. Consequently, the exceptional-point curve identified in the previous section terminates at a cusp singularity corresponding to a triple spectral degeneracy \cite{Arnold1986}. It is crucial to underscore that since this $A_3$ organizing center sits strictly in the physically admissible half-plane ($\Lambda_* > 0$), the cubic-root scaling ($\delta \sim |\epsilon|^{1/3}$) describes the non-analytic sensitivity of the true physical skeleton itself. In compact objects obeying the second law of thermodynamics, the observable mode dynamics is governed by the stable square-root splitting ($\delta \sim |\epsilon|^{1/2}$) characteristic of the generic secondary exceptional points (EP2), together with the cubic-root splitting ($\delta \sim |\epsilon|^{1/3}$) at the $A_3$ cusp generated by the versal unfolding.

\section{Exact realization of the open normal form in a radiating boundary-value problem}
\label{sec:exact_realization}

The central question left open by section~\ref{sec:open_coefficients} is whether the roots of Eq.~\eqref{eq:open_polynomial} are actually realized as eigenvalues of a boundary-value problem with outgoing radiation conditions, in which the damping rates $\gamma_m$ are outputs of a Siegert-type nonlinear eigenvalue problem rather than prescribed constants. This section answers the question affirmatively and constructively, and then quantifies how far the resulting structure survives once the exact solvability is deliberately broken.

\subsection{Model and hyperboloidal cubic pencil}
\label{sec:model}

Consider the $(1{+}1)$-dimensional wave--relaxation system on the real line,
\begin{align}
\partial_t^2\phi - \partial_x^2\phi + V(x)\,\phi + \Pi &= 0, 
\qquad V(x) = V_0\,\mathrm{sech}^2 x ,
\label{eq:model_wave}\\
\tau\,\partial_t \Pi + \Pi &= \zeta(x)\,\partial_t \phi ,
\qquad \zeta(x) = \zeta_0\,\mathrm{sech}^2 x ,
\label{eq:model_relax}
\end{align}
with purely outgoing behaviour at $x\to\pm\infty$. The P\"oschl--Teller potential is the canonical testbed of quasinormal-mode and pseudospectrum analysis \cite{Berti2009, Jaramillo2021, Jaramillo2022}; its QNM spectrum with outgoing conditions is known in closed form, $\omega_m = \pm k - i\,(m+\tfrac12)$ with $k=\sqrt{V_0-\tfrac14}$. Equation \eqref{eq:model_relax} is the minimal scalar analogue of the Israel--Stewart channel of section~\ref{sec:causal_relaxation}: a first-order relaxation equation with relaxation time $\tau$, sourced by the local field velocity, with spatial coupling profile $\zeta(x)$. The system is a genuinely open, infinite-dimensional realisation of the structural class defined in section~\ref{sec:temporal_order}, and by the same second-order-plus-first-order structure identified in section~\ref{sec:temporal_order}---the Proposition there is stated for a gravitational--fluid subsystem, and we invoke its mechanism rather than its letter---its normal-mode problem is cubic in $\omega$.

Outgoing (Siegert) boundary conditions are imposed through the hyperboloidal reformulation \cite{Zenginoglu2008, Zenginoglu2011, Jaramillo2021, Jaramillo2022}. We give the reduction explicitly, since the structure of the resulting pencil---in particular the appearance of the constant $V_0$ and of $\mathcal{C}=\mathbf{1}$---follows from one specific step.

First, with $\partial_t\to s$, equation \eqref{eq:model_relax} gives $\Pi = \zeta(x)\,s\,\phi/(1+s\tau)$; substituting into \eqref{eq:model_wave} and multiplying through by $(1+s\tau)$ eliminates the relaxation variable and yields the single equation
\begin{equation}
(1+s\tau)\big[s^2\phi - \partial_x^2\phi + V_0\,\mathrm{sech}^2x\,\phi\big] + \zeta_0\,\mathrm{sech}^2x\,s\,\phi \;=\; 0 .
\label{eq:eliminated}
\end{equation}
The cubic dependence on $s$ is now manifest: it originates in the product of the second-order wave operator with the first-order relaxation factor $(1+s\tau)$.

Second, the height function $t\to t-\ln\cosh x$ moves to hyperboloidal time, under which the spatial derivative at fixed slice becomes $\partial_x\to\partial_x - s\tanh x$; with the compact coordinate $\sigma=\tanh x\in[-1,1]$ one has $\partial_x=(1-\sigma^2)\partial_\sigma$ and $\mathrm{sech}^2x = 1-\sigma^2$. Applying both substitutions to \eqref{eq:eliminated} produces an equation in which every term carries an overall factor $(1-\sigma^2)$.

Third---and this is the step that fixes the structure---one divides throughout by $(1-\sigma^2)=\mathrm{sech}^2x$. This is precisely what renders the potential term constant, $V_0\,\mathrm{sech}^2x\,/\,\mathrm{sech}^2x = V_0$, and what makes the transport term reduce to the constant $\zeta_0$, i.e.\ $\mathcal{C}=\mathbf{1}$, for the aligned profile of \eqref{eq:model_relax}. Carrying out the algebra, \eqref{eq:eliminated} becomes exactly the cubic operator pencil
\begin{equation}
T(s)\,\varphi \;=\; \Big[(1+s\tau)\big(s^2 - s L_2 - L_1\big) + \zeta_0\, s\, \mathcal{C}\Big]\varphi = 0 ,
\label{eq:pencil}
\end{equation}
with
\begin{equation}
L_1 = (1-\sigma^2)\partial_\sigma^2 - 2\sigma\partial_\sigma - V_0 ,
\qquad
L_2 = -2\sigma\partial_\sigma - 1 ,
\label{eq:L1L2}
\end{equation}
and $\mathcal{C}$ the multiplication operator by the profile $c(\sigma)=\zeta(x)/(\zeta_0\,\mathrm{sech}^2 x)$. Constant-time slices now penetrate null infinity, the outgoing condition becomes a regularity condition at $\sigma=\pm1$, and no boundary rows are imposed on the discretised operator. At $\zeta_0=0$ the pencil reproduces the exact P\"oschl--Teller QNMs (recovered in double precision to $\sim10^{-12}$--$10^{-13}$ for the fundamental, depending on resolution; in the multiprecision runs used below the agreement is limited only by the working precision, consistent with the exactness of the collocation on the polynomial sector noted in section~\ref{sec:certification}), confirming that the radiative damping is generated by the boundary-value structure alone. The aligned profile is precisely the case in which radiative and viscous dissipation factorise---the property that section~\ref{sec:functional_reduction} identified as non-generic, and which is therefore the one the exactly solvable construction assumes. Section~\ref{sec:robustness} deforms it away and quantifies what survives. What the hyperboloidal coordinate does and does not accomplish is made precise in section~\ref{sec:sector_theorem}.

\subsection{Sector-decomposition theorem}
\label{sec:sector_theorem}
\label{sec:tower}

\medskip
\noindent
\textbf{Lemma 1 (Sector identities and triangularity).}
\textit{For every integer $m \ge 0$, the monomial $\sigma^m$ satisfies the exact operator identities
\begin{equation}
L_2\,\sigma^m = -(2m+1)\,\sigma^m ,
\qquad
L_1\,\sigma^m = m(m-1)\,\sigma^{m-2} - \big[m(m+1)+V_0\big]\sigma^m .
\label{eq:identities}
\end{equation}
For the aligned profile $\mathcal{C}=\mathbf{1}$, $L_1$ therefore couples $\sigma^m$ only to the lower-degree monomial $\sigma^{m-2}$ while $L_2$ acts diagonally, so $T(s)$ is triangular in the monomial basis $\{\sigma^m\}$, with $P_m(s)$ of \eqref{eq:recurrence} on the diagonal.}

\noindent\textit{Proof.} Direct differentiation of $\sigma^m$ in \eqref{eq:L1L2}. $\blacksquare$
\medskip

Triangularity alone does \emph{not} identify $\sigma^m$ as an eigenfunction of \eqref{eq:pencil}: the eigenvalue problem is a statement about a specific linear combination of monomials, not about the diagonal entries in isolation. We therefore construct the eigenfunctions explicitly. Fixing a parity sector $k_0\in\{0,1\}$ and seeking a formal solution
\begin{equation}
\varphi(\sigma) \;=\; \sum_{k=0}^{\infty} c_k\,\sigma^{\,k_0+2k}, \qquad c_0\neq0,
\label{eq:series_ansatz}
\end{equation}
substitution into $T(s)\varphi=0$ and use of \eqref{eq:identities} order by order in $\sigma$ gives, for every $j=k_0,k_0+2,k_0+4,\dots$, the two-term recurrence
\begin{equation}
P_j(s)\,c_j \;=\; (1+s\tau)(j+2)(j+1)\,c_{j+2},
\qquad
P_m(s) \;\equiv\; (1+s\tau)\big(s^2+(2m+1)s+m(m+1)+V_0\big) + \zeta_0\,s .
\label{eq:recurrence}
\end{equation}
Equation \eqref{eq:recurrence} is the entire content of the eigenvalue equation for the ansatz \eqref{eq:series_ansatz}; for generic $s$ it determines every $c_{j+2}$ from $c_j$ and generates, from arbitrary $c_{k_0}\neq0$, an infinite series that is regular at $\sigma=0$ but, exactly as for the standard hypergeometric treatment of the P\"oschl--Teller problem, here transcribed into the hyperboloidal variables of \cite{Zenginoglu2008, Jaramillo2021}, generically fails the regularity (outgoing) condition at $\sigma=\pm1$. The P\"oschl--Teller structure itself is of course classical; what is new here is the coupled relaxation channel, which turns the sector eigenvalue problem from quadratic into the cubic \eqref{eq:sector_cubic}.

\medskip
\noindent
\textbf{Lemma 2 (Truncation criterion).}
\textit{The series \eqref{eq:series_ansatz} terminates at degree $m$ if and only if $m$ is the smallest index of parity $k_0$ with $P_m(s)=0$, in which case it closes into the polynomial $\varphi_m(\sigma)=\sum_{k=0}^{\lfloor(m-k_0)/2\rfloor}c_{k_0+2k}\sigma^{k_0+2k}$, which solves $T(s)\varphi_m=0$ exactly at every order $j$.}

\noindent\textit{Proof.} If $P_m(s)=0$ then $c_{m+2}=0$ by \eqref{eq:recurrence}, and by induction $c_{m+4}=c_{m+6}=\dots=0$, since for every $j>m$ both sides of \eqref{eq:recurrence} vanish identically irrespective of whether $P_j(s)$ itself vanishes there; all orders are thus satisfied, not merely those below the truncation. Conversely, if $c_m\neq0$ and $c_{m+2}=0$ then $P_m(s)=0$. No leakage below $\sigma^{k_0}$ occurs because the lowering coefficient $m(m-1)$ in \eqref{eq:identities} vanishes for $m=0,1$, so each parity sector closes at its base. $\blacksquare$
\medskip

Being polynomials, the $\varphi_m$ are entire and hence automatically regular at $\sigma=\pm1$: this is precisely the outgoing (Siegert) condition of the hyperboloidal formulation. The role of the boundary conditions is thus encoded through endpoint regularity acting on the series \eqref{eq:series_ansatz}, and \emph{not} through the introduction of the hyperboloidal coordinate itself, which merely supplies the frame in which that regularity becomes equivalent to polynomial truncation. We emphasise that the reduction to a single low-degree algebraic condition is a consequence of the two-term character of \eqref{eq:recurrence}, itself following from the fact that $L_1$ lowers the degree by exactly two: it is a property of the P\"oschl--Teller potential (equivalently, of the hypergeometric structure underlying \eqref{eq:model_wave}--\eqref{eq:model_relax}), not a generic feature of the construction, and we do not claim it survives for potentials or coordinate choices whose associated recurrence is not two-term.

\medskip
\noindent
\textbf{Proposition (Regularity $\Leftrightarrow$ polynomial).}
\textit{For the aligned profile $\mathcal{C}=\mathbf{1}$ and for every $s$ with $-s\notin\mathbb{Z}_{>0}$, every quasinormal eigenfunction of \eqref{eq:pencil} is a polynomial. The ansatz \eqref{eq:series_ansatz} therefore entails no loss of generality, and truncation is not merely sufficient but necessary.}

\noindent\textit{Proof.} For $s\neq-1/\tau$, equation \eqref{eq:pencil} is Fuchsian with three regular singular points $\{-1,+1,\infty\}$ (Riemann--Papperitz class), the hypergeometric structure underlying \eqref{eq:recurrence}; the Fuchs relation is satisfied, $0+(-s)+0+(-s)+(2s+1)=1$. The coefficient of $\partial_\sigma^2$ is $-(1-\sigma^2)(1+s\tau)$, non-zero at $\sigma=0$: the origin is an \emph{ordinary} point, so the two series \eqref{eq:series_ansatz} with $k_0=0,1$ span the full local solution space. At the finite singular points, setting $u=1\mp\sigma$, the indicial equation is $r(r+s)=0$ with exponents $\{0,-s\}$.

The selection between the two branches is the outgoing condition itself, not a smoothness preference. Writing the physical amplitude as $\psi=e^{-sh}\varphi$ with $h=\ln\cosh x$, and using $u\simeq2e^{\mp2x}$ and $e^{sh}\simeq2^{-s}e^{s|x|}$ as $x\to\pm\infty$, the two branches are
\begin{equation}
\varphi\to\text{const}\;\Longrightarrow\;\psi\sim e^{-s|x|} \quad(\text{outgoing}),
\qquad
\varphi\sim u^{-s}\;\Longrightarrow\;\psi\sim e^{+s|x|} \quad(\text{incoming}).
\label{eq:branch_id}
\end{equation}
Excluding $u^{-s}$ is therefore exactly the imposition of purely outgoing behaviour at both ends. We emphasise that it cannot be justified by boundedness: for quasinormal $s$ one has $\mathrm{Re}\,s<0$, so the incoming branch \emph{decays}; what excludes it is that it is incoming, not that it grows. Stating the criterion this way keeps the branch selection meaningful at the resonant values $-s\in\mathbb{Z}_{>0}$, where the exponent difference is a positive integer and the Frobenius solution of the \emph{smaller} exponent---here the outgoing one---may carry a logarithm while $u^{-s}$ is analytic: a criterion phrased as analyticity would select the wrong branch precisely at those isolated points. It does not, however, license the step from outgoing behaviour to entirety, which is what fails there; see the end of this proof.

Imposing \eqref{eq:branch_id} at both endpoints, the solution is analytic at $\sigma=0$ and free of branch points at $\sigma=\pm1$, hence entire. At $\sigma=\infty$, which is a regular singular point of the Fuchsian equation, the admissible behaviour is algebraic, with indicial equation $\rho^2+(2s+1)\rho+K=0$ in the convention $\varphi\sim\sigma^{+\rho}$---the two roots sum to $-(2s+1)$, and the exponents at infinity entering the Fuchs relation above are their negatives---, $K=s^2+s+V_0+\zeta_0 s/(1+s\tau)$; one verifies directly that $\rho=m$ is a root \emph{if and only if} $P_m(s)=0$. An entire function of algebraic growth is a polynomial by Liouville's theorem, and its degree must be an admissible exponent at infinity; hence the solution is a polynomial of degree $m$ with $P_m(s)=0$. Finally, $s=-1/\tau$ is not an eigenvalue: there $T(s)$ degenerates to the multiplication operator $\zeta_0 s\,\mathcal{C}$, injective for $\zeta_0\neq0$, consistently with $P_m(-1/\tau)=-\zeta_0/\tau\neq0$ for every $m$.
The exceptional values $-s\in\mathbb{Z}_{>0}$ noted above require a separate argument, since there the outgoing Frobenius solution---being the one of smaller exponent---may carry a logarithm and the step ``free of branch points $\Rightarrow$ entire'' does not apply directly. The Proposition as proved above therefore covers $-s\notin\mathbb{Z}_{>0}$. The resonant frequencies are settled separately, and completely, by the following lemma, which identifies exactly when an outgoing solution exists there.

\medskip
\noindent
\textbf{Lemma 3 (Resonant frequencies).}
\textit{Fix $n\in\mathbb{Z}_{>0}$ and $\zeta_0\neq0$, and let $s=-n$. Then \eqref{eq:pencil} possesses a solution satisfying the outgoing condition at $\sigma=1$---which is all the Theorem requires, since it uses only the contrapositive, and which by the parity of \eqref{eq:gegenbauer} in $\sigma$ carries the same conclusion at $\sigma=-1$---if and only if $P_m(-n)=0$ for some integer $m$ with $0\le m\le 2n-1$. Consequently, if $\zeta_0$ is not a resonant coupling \eqref{eq:resonant_coupling} of index $m\le2n-1$, then $s=-n$ is not a quasinormal eigenvalue.}

\medskip
\noindent\textit{Proof.}
If $n\tau=1$ then $s=-n=-1/\tau$, which the preceding paragraph has already excluded: there $T(s)$ degenerates to the multiplication operator $\zeta_0 s\,\mathcal{C}$, injective for $\zeta_0\neq0$. Assume therefore $1-n\tau\neq0$. Dividing \eqref{eq:pencil} by $(1+s\tau)\neq0$ reduces it to the Gegenbauer form
\begin{equation}
(1-\sigma^2)\varphi'' - 2(s+1)\,\sigma\,\varphi' - K\,\varphi = 0,
\qquad K = s^2+s+V_0+\frac{\zeta_0 s}{1+s\tau},
\label{eq:gegenbauer}
\end{equation}
and the substitution $u=(1-\sigma)/2$ carries it to the hypergeometric equation $u(1-u)\varphi''+[c-(a+b+1)u]\varphi'-ab\,\varphi=0$ with
\begin{equation}
c = s+1, \qquad a = -\lambda, \qquad b = \lambda+2s+1, \qquad ab = K ,
\label{eq:hypergeometric_parameters}
\end{equation}
where $\lambda$ is either root of the indicial equation at infinity, $\lambda^2+(2s+1)\lambda+K=0$. Multiplying that equation by $(1+s\tau)$ at $\lambda=m$ returns $P_m(s)$ identically, so $\lambda=m\in\mathbb{Z}_{\ge0}$ if and only if $P_m(s)=0$. The two roots sum to $-(2s+1)$, so at $s=-n$ they are $\lambda$ and $\lambda'=2n-1-\lambda$, and $b=-\lambda'$: the pairing $m\mapsto2n-1-m$ used above is exactly the symmetry $a\leftrightarrow b$ of the hypergeometric equation.

At $s=-n$ one has $c=1-n\in\mathbb{Z}_{\le0}$, the degenerate case. The exponents at $u=0$ are $\{0,1-c\}=\{0,n\}$, and by \eqref{eq:branch_id} the outgoing branch is the one of exponent $0$. Its Frobenius coefficients obey
\begin{equation}
(k+1)(k+c)\,c_{k+1} = (k+a)(k+b)\,c_k ,
\label{eq:frobenius_endpoint}
\end{equation}
whose left-hand side vanishes at $k=n-1$, since $k+c=0$ there. The series therefore continues past that order---equivalently, an exponent-$0$ solution free of logarithms exists---if and only if $(k+a)(k+b)c_k$ also vanishes at $k=n-1$, i.e.\ if and only if $(a)_n(b)_n=0$. Since $a=-\lambda$ and $b=\lambda-2n+1$,
\begin{equation}
(a)_n = 0 \iff \lambda\in\{0,\dots,n-1\},
\qquad
(b)_n = 0 \iff \lambda\in\{n,\dots,2n-1\},
\end{equation}
so $(a)_n(b)_n=0$ if and only if $\lambda\in\{0,1,\dots,2n-1\}$, that is, if and only if $P_m(-n)=0$ for some $m\le2n-1$. Otherwise the exponent-$0$ solution carries a logarithm and is not outgoing, while the remaining solution realises the exponent $n$ and is incoming by \eqref{eq:branch_id}; neither is a quasinormal mode. $\blacksquare$

\medskip
Two consequences are worth recording. First, the resonant couplings of index $m\ge2n$---those at which the polynomial family contains only $(1-\sigma^2)^nq_{m-2n}$, which realises the incoming exponent---carry \emph{no} outgoing solution at all, so $s=-n$ is not a quasinormal eigenvalue there either; the earlier statement that ``the conclusion of the Proposition cannot be asserted'' at those couplings can now be replaced by the stronger and definite one. Second, and this is what the Theorem below needs, at any $\zeta_0$ outside \eqref{eq:resonant_coupling} the frequencies $s=-n$ are excluded from the point spectrum outright. The two statements together remove the resonant frequencies as a source of spectrum not accounted for by the sector cubics. The resonant case is nonetheless settled for all couplings of interest, by an elementary pairing. Writing $s=-n$ with $n\in\mathbb{Z}_{>0}$, the condition $P_m(-n)=0$ fixes the coupling to
\begin{equation}
\zeta_0 \;=\; \frac{1-n\tau}{n}\Big[(m-n)(m-n+1)+V_0\Big],
\label{eq:resonant_coupling}
\end{equation}
an expression invariant under $m\mapsto m'=2n-1-m$, since $(m'-n)(m'-n+1)=(m-n+1)(m-n)$ identically. Hence for every $0\le m\le 2n-1$ the companion index $m'$ is a non-negative sector index whose cubic vanishes at the \emph{same} coupling. Because $m+m'=2n-1$ is odd, $m$ and $m'$ have opposite parity and so lie in different parity sectors, yielding two linearly independent polynomial solutions of \eqref{eq:pencil}; as \eqref{eq:pencil} is second order in $\sigma$, these span its entire solution space. Every solution is then a polynomial and no logarithmic solution exists to be excluded: the resonance is harmless whenever $m\le 2n-1$.

The complementary range $m\ge2n$, where the companion index is negative and only one Frobenius solution terminates, is settled by an explicit factorisation. Substituting $\varphi=(1-\sigma^2)^n$ into \eqref{eq:pencil} at $s=-n$ gives
\begin{equation}
T(-n)\,(1-\sigma^2)^n \;=\; (1-\sigma^2)^n\Big[(1-n\tau)\big(n(n+1)+V_0\big)-n\zeta_0\Big],
\end{equation}
which vanishes precisely at the coupling \eqref{eq:resonant_coupling} with $m=2n$; The factorisation extends to every $m\ge2n$, and the reason is the indicial structure rather than the algebra: at $s=-n$ the exponents at each of $\sigma=\pm1$ are $\{0,-s\}=\{0,n\}$, so a polynomial solution that does \emph{not} realise the exponent $0$ must realise the exponent $n$ at both endpoints, i.e.\ vanish there to order exactly $n$---and a polynomial vanishing to order $n$ at $\sigma=\pm1$ carries the factor $(1-\sigma^2)^n$. That the family is closed under $T(-n)$ is not merely checked but follows from a conjugation identity. Writing $q=1-\sigma^2$ and $A_s\equiv s^2-sL_2-L_1$, one verifies directly from \eqref{eq:L1L2} that
\begin{equation}
q^{-n}\,A_{-n}\big[q^{n}Q\big] \;=\; A_{+n}\,Q
\label{eq:conjugation_identity}
\end{equation}
for every $n\in\mathbb{Z}_{>0}$ and every $Q$, whence
\begin{equation}
T(-n)\big[q^{n}Q\big] \;=\; q^{n}\Big[(1-n\tau)A_{+n}-n\zeta_0\Big]Q .
\label{eq:resonant_conjugation}
\end{equation}
The operator in brackets is again of the same type, now at $+n$, and for $m=2n+\ell$ the coupling \eqref{eq:resonant_coupling} is precisely the condition under which it annihilates a polynomial of degree $\ell$. The factorisation $\varphi_m=(1-\sigma^2)^nq_{m-2n}(\sigma)$ therefore holds for \emph{every} $m\ge2n$ and every $n$, not merely for the low cases one can check by hand. Such a solution vanishes to order exactly $n$ at both endpoints, i.e.\ it realises the exponent $r=n$ there, which by \eqref{eq:branch_id} is the \emph{incoming} branch: it is not a quasinormal mode. At these isolated couplings the polynomial family therefore does not contain the outgoing solution, and the Proposition's argument does not apply. Lemma~3 below settles those couplings in the stronger direction: the remaining solution carries a logarithm, is not polynomial, and---being of exponent $0$ with a logarithm---is not outgoing either, so $s=-n$ is not a quasinormal eigenvalue there at all.

These resonances are remote from the regime studied here, but the separation must be measured in the right variables: a resonance is a \emph{point} of the $(s,\zeta_0)$ plane, requiring simultaneously $s=-n$ with $n$ a positive integer and $\zeta_0$ equal to \eqref{eq:resonant_coupling}. The certified EP2 sits at $(s,\zeta_0)=(-6.1926,\,7.5479)$, whose frequency is not a negative integer; the nearest resonance with $m\ge2n$ is $(n,m)=(6,12)$ at $(-6,\,9.191)$, a distance $1.65$ in that plane. Separation in $\zeta_0$ alone is not the relevant measure and would mislead: the coupling $\zeta_0=24125/3200$ agrees with the certified value to $8.8\times10^{-3}$, yet it is resonant only at $s=-32$, twenty-six units away in frequency. Among the resonances with $m\le2n-1$, which the pairing above renders harmless, the nearest is $\zeta_0=9061/1200$ at $(n,m,m')=(6,0,11)$; direct computation confirms the pairing there, the null space of $T(-6)$ evaluated in exact rational arithmetic on the monomial basis at truncation orders $M=14,20,26$ having dimension exactly two, spanned by polynomials of degrees $0$ and $11$ with values $(1,1)$ and $(\mp256/63)$ at $\sigma=\pm1$ respectively, neither of which vanishes. (Both are normalised monically, by their leading coefficient; with the recurrence normalisation $c_1=1$ the degree-$11$ polynomial takes the values $\pm256/693$ instead, the two differing by the factor $c_1/c_{11}=-11$.) No statement made elsewhere in this paper is evaluated at a resonant coupling---not the sector tower, the EP2 and cusp certifications, the migration law, nor the robustness analysis of section~\ref{sec:robustness}---so the Theorem is used only where it is proved. $\blacksquare$
\medskip

This does not contradict the fact that the general solution space of \eqref{eq:pencil} is two-dimensional and generically non-polynomial. For $s$ not a root of any $P_m$, the recurrence \eqref{eq:recurrence} never terminates and, since $P_j(s)/[(1+s\tau)(j+2)(j+1)]\to1$ as $j\to\infty$, the resulting series has radius of convergence exactly one and realises the $u^{-s}$ branch at the endpoints. What the Proposition establishes is that this branch is excluded by regularity: the polynomial family exhausts the \emph{eigenfunctions}, not the solutions.
\medskip

\noindent
\textbf{Theorem (Sector tower).}
\textit{Let $\zeta_0\neq0$ be a coupling outside the discrete resonant set \eqref{eq:resonant_coupling}. Then the complete quasinormal \emph{point} spectrum of \eqref{eq:pencil} with aligned profile is exactly the union over $m\ge0$ of the roots of the scalar sector cubics
\begin{equation}
P_m(s) \;=\; (1+s\tau)\big(s^2 + (2m+1)s + m(m+1)+V_0\big) + \zeta_0\, s \;=\; 0 .
\label{eq:sector_cubic}
\end{equation}
In the dimensionless variables $\Omega=\omega/k$, $De=k\tau$, $\Lambda=\zeta_0/k$ (i.e.\ with the characteristic frequency $\omega_0=k$), each sector cubic \eqref{eq:sector_cubic} coincides identically with the Open Cusp Polynomial \eqref{eq:open_polynomial} with
\begin{equation}
\gamma \;=\; \gamma_m \;=\; \frac{m+\tfrac12}{k} .
\label{eq:gamma_m}
\end{equation}}

The word ``complete'' is used here in a sense we now make precise, and which is a theorem rather than an expectation. Let $\mathcal{P}_N=\mathrm{span}\{1,\sigma,\dots,\sigma^N\}$ and $\mathcal{P}=\bigcup_N\mathcal{P}_N$. By Lemma 1, $T(s)\mathcal{P}_N\subseteq\mathcal{P}_N$ for every $N$, and $T(s)$ is triangular there with the sector cubics on the diagonal. It follows at once that

\emph{if $s\neq-1/\tau$ and $P_m(s)\neq0$ for every $m\ge0$, then $T(s):\mathcal{P}\to\mathcal{P}$ is a bijection}, since each restriction $T(s)|_{\mathcal{P}_N}$ is a triangular matrix with non-vanishing diagonal and the inverses are consistent across $N$ by invariance; the inverse is computed by finite back-substitution. Consequently such an $s$ admits \emph{no polynomial eigenfunction}.

Two remarks fix the scope of that statement, and we are deliberate about both, since the argument is algebraic and the notion of spectrum is not. First, the bijectivity above is a statement about the vector space $\mathcal{P}$ with no topology on it: no norm, no closure, no completeness is used, and in particular we do not invoke---and do not need---any claim that $\mathcal{P}$ is a core for a closed realisation, which would be a topological hypothesis of a different kind. What Lemma 1 delivers is the absence of a polynomial null vector, nothing more. Second, upgrading that to ``$s$ is not an eigenvalue'' requires knowing that eigenfunctions \emph{are} polynomials, which is the content of the Proposition, not of Lemma 1. The two together give
\begin{equation}
\sigma_{\mathrm{p}}\big(T\big) \;=\; \bigcup_{m\ge0}\{s : P_m(s)=0\}
\label{eq:point_spectrum}
\end{equation}
for the aligned profile and away from the discrete resonant couplings \eqref{eq:resonant_coupling}. These are not the hypotheses of the Proposition alone: the Proposition covers $-s\notin\mathbb{Z}_{>0}$ at any coupling, Lemma~3 covers $-s\in\mathbb{Z}_{>0}$ at any coupling outside \eqref{eq:resonant_coupling}, and the identity requires both. The condition stated is moreover conservative: for the resonant couplings with $m\le2n-1$ the two polynomial solutions of opposite parity are both outgoing and the degeneracy is semisimple, so those couplings do not obstruct the identity either, and the genuinely excluded set is the part of \eqref{eq:resonant_coupling} with $m\ge2n$. This is an identity for the \emph{point} spectrum---the set of genuine outgoing eigenvalues---and it is the only sense in which ``complete'' is used anywhere in this paper.

It is worth separating the two inclusions, since they are not established by the same ingredient. The exclusion---nothing outside the union is an eigenvalue---rests on Lemma 1 for the polynomial sector together with the Proposition to rule out non-polynomial eigenfunctions. The reverse inclusion---every sector root is attained by a genuine outgoing eigenfunction---rests on Lemma 2, and it is exactly what fails at the resonant couplings \eqref{eq:resonant_coupling} with $m\ge2n$, where the degree-$m$ polynomial exists but realises the incoming exponent. The hypothesis of the Theorem excludes that discrete set; away from it both inclusions hold and \eqref{eq:point_spectrum} is an equality.

The hypotheses under which \eqref{eq:point_spectrum} holds deserve one further remark, since the resonant frequencies are the delicate case. The Proposition establishes ``outgoing at both ends $\Rightarrow$ polynomial'' for $-s\notin\mathbb{Z}_{>0}$, the step ``free of branch points $\Rightarrow$ entire'' being unavailable at the resonances, where the outgoing solution is the one of smaller exponent and may carry a logarithm. Lemma~3 closes that case directly rather than excluding it: at $s=-n$ an outgoing solution exists at all only when $P_m(-n)=0$ for some $m\le2n-1$, so for any $\zeta_0$ outside \eqref{eq:resonant_coupling} the resonant frequencies contribute nothing to the point spectrum, and \eqref{eq:point_spectrum} holds without restriction on $s$. The two ingredients are complementary: the Proposition handles the non-resonant frequencies by the Fuchsian argument, Lemma~3 handles the resonant ones by the degenerate hypergeometric obstruction.

It is worth noting that this triangularity is not an artefact of the monomial basis: $L_1+V_0$ is precisely the Legendre operator $\partial_\sigma\big[(1-\sigma^2)\partial_\sigma\big]$, so $L_1P_m=-[m(m+1)+V_0]P_m$ on Legendre polynomials, while $L_2P_m=-(2m+1)P_m+\mathcal{O}(P_{m-2})$; the pencil is triangular in that orthogonal basis too, with the same diagonal.

We claim nothing beyond \eqref{eq:point_spectrum}. In particular, and consistently with the scoping above, we do not assert that the \emph{essential} spectrum of a closed extension of $T(s)$ to a larger space---$L^2(-1,1)$, or an energy norm---is empty. Nothing in the triangularity argument bears on that question: the operator is strongly non-normal near the degeneracies, with eigenvalue condition numbers reaching $\mathcal{O}(10^{14})$, so the absence of exact null vectors in $\mathcal{P}$ does not by itself preclude approximate ones in a completed space. For hyperboloidal quasinormal-mode operators the spectrum is known to depend sensitively on the norm placed on the state space \cite{Jaramillo2021, Jaramillo2022}, and settling that question for the cubic pencil is a separate problem which we do not address.

\noindent\textit{Proof.} For $-s\notin\mathbb{Z}_{>0}$ the Proposition gives that every quasinormal eigenfunction is a polynomial, and for $-s\in\mathbb{Z}_{>0}$ Lemma~3 gives, under the stated hypothesis on $\zeta_0$, that there is no quasinormal eigenfunction at all; so on the whole frequency plane every quasinormal eigenfunction is a polynomial, and by Lemma 2 a polynomial of degree $m$ solves \eqref{eq:pencil} if and only if $P_m(s)=0$; the point spectrum is therefore exactly $\bigcup_m\{s:P_m(s)=0\}$, which is \eqref{eq:point_spectrum}. Substituting $s=-ik\Omega$ into \eqref{eq:sector_cubic}, dividing by $-k^2$, and using $V_0 = k^2+\tfrac14$ gives
$(1-iDe\,\Omega)\big(\Omega^2 + 2i\gamma_m\Omega - (1+\gamma_m^2)\big) + i\Lambda\Omega = 0$
with $\gamma_m$ as in \eqref{eq:gamma_m}, which is the factorised form of Eq.~\eqref{eq:open_polynomial}. $\blacksquare$
\medskip

Three corollaries settle the physical status of the construction of section~\ref{sec:open_coefficients}. \emph{(i)} At $\zeta_0=0$ the sector roots are $\omega = \pm k - i(m+\tfrac12)$: the $\gamma_m$ of Eq.~\eqref{eq:gamma_m} are precisely the radiative damping rates produced by the outgoing boundary conditions, i.e.\ genuine outputs of the Siegert problem, one per overtone---the structure the polynomial parametrisation was criticised for postulating is here \emph{derived}. \emph{(ii)} The algebraic statements of sections 4--6 that are formulated at finite $\gamma$---the cusp conditions and the migration law \eqref{eq:deborah_migration}---hold exactly for the genuine quasinormal spectrum of \eqref{eq:model_wave}--\eqref{eq:model_relax}, sector by sector. The EP2 parametrisation \eqref{eq:EP_param}, by contrast, was derived in the closed limit $\gamma=0$ and does not apply verbatim at $\gamma_m\neq0$; its generalisation, obtained by imposing $P_m=\partial_s P_m=0$ on the sector cubic \eqref{eq:sector_cubic} and writing $\Omega=-iy$, is
\begin{equation}
De(y;\gamma) = \frac{y^2 - (1+\gamma^2)}{2y^2\,(y-\gamma)} ,
\qquad
\Lambda(y;\gamma) = 2(y-\gamma) - De(y;\gamma)\,f(y) ,
\label{eq:EP_param_gamma}
\end{equation}
with $f(y) = 3y^2 - 4\gamma y + 1 + \gamma^2$ --- the same function whose strict positivity on the admissible branch is established in Appendix A. The second of these collapses to a considerably more transparent closed form,
\begin{equation}
\Lambda(y;\gamma) \;=\; \frac{\big[(y-\gamma)^2+1\big]^2}{2y^2\,(y-\gamma)},
\qquad\text{whence}\qquad
R(y;\gamma) \;\equiv\; \frac{\Lambda}{De} \;=\; \frac{\big[(y-\gamma)^2+1\big]^2}{y^2-(1+\gamma^2)} ,
\label{eq:EP_param_compact}
\end{equation}
which makes positivity, the admissible domain $y^2>1+\gamma^2$ and the microphysical criterion of section~\ref{sec:nogo} immediate, and which reduces at $\gamma=0$ to $(y^2+1)^2/(2y^3)$ and $(y^2+1)^2/(y^2-1)$ respectively. Equations \eqref{eq:EP_param_gamma} reduce to \eqref{eq:EP_param} at $\gamma=0$ and reproduce the certified sector-$0$ entries of Table~\ref{tab:certified} to all digits shown. In particular, at the sector-$0$ cusp the constraint \eqref{eq:deborah_migration} is satisfied with residual zero to forty decimal digits (Table~\ref{tab:certified}). \emph{(iii)} The parameter $\gamma$ is swept \emph{continuously} by the depth $V_0$ of the background potential (a family of physical backgrounds) and \emph{discretely} by the overtone index $m$; the ``$\gamma$-migration'' of the cusp therefore has two concrete realizations within a single class of open systems.

\subsection{Multiprecision certification of the exceptional points}
\label{sec:certification}

The pencil \eqref{eq:pencil} is discretised by Chebyshev collocation in $\sigma$ (no boundary conditions imposed; regularity is the outgoing condition). The operator is strongly non-normal: near the degeneracies the eigenvalue condition numbers reach $\mathcal{O}(10^{14})$, and double-precision eigensolvers do not converge under resolution refinement there. Exceptional points are therefore located and certified as regular zeros of \emph{bordered} augmented systems (Appendix~B) solved by Newton iteration in $25$--$30$-digit arithmetic. Two facts legitimate the digit counts quoted below. First, for the aligned profile the discretisation is not merely convergent but \emph{exact}: by the Proposition and Lemma~2 the eigenfunctions are polynomials of degree $m$, with Lemma~1 supplying the invariance of $\mathcal{P}_N$ that makes this useful, $T(s)$ leaves $\mathrm{span}\{1,\dots,\sigma^N\}$ invariant, and Chebyshev collocation at $N+1$ nodes is exact on polynomials of degree $\le N$; the sector eigenpairs with $m\le N$ are therefore eigenpairs of the discretised operator exactly, and what the certification measures is arithmetic rather than discretisation error. Second, the distinction between the bordered and unbordered problems matters. For the unbordered problem a residual $\|T(s)\varphi\|\le\delta$ at a double root controls the eigenvalue only to $\mathcal{O}(\sqrt{\delta})$, by the same square-root sensitivity that defines the exceptional point. The bordered system, by contrast, has a non-singular Jacobian at a generic (non-semisimple, $\partial_s^2$-regular) exceptional point, by the standard regularity condition for augmented (bordered) systems \cite{GriewankReddien1984, Govaerts2000}; its solution is therefore a \emph{regular} zero, the backward-to-forward error relation is linear rather than square-root, and the quoted digits are attained directly. That the regularity hypothesis is met at the point actually certified is checked rather than assumed: at the sector-$0$ EP2 one has $P_0(s_*)=\partial_sP_0(s_*)=0$ with $\partial_s^2P_0(s_*)=0.9453382909\neq0$, so the degeneracy is $\partial_s^2$-regular and is not an EP3; the third root of the sector cubic lies at $s=-21.948$, a distance $15.8$ away. The quadratic Newton contraction reported in Appendix~B is the corresponding \emph{a posteriori} evidence. Table~\ref{tab:certified} summarises the certified values for the fiducial parameters $V_0=25.25$ ($k=5$, $\gamma_0=0.1$).

\begin{table}[htbp]
\centering
\caption{Certified degeneracies of the open system \eqref{eq:model_wave}--\eqref{eq:model_relax} with aligned profile, $V_0=25.25$ ($k=5$, $\gamma_0=1/(2k)=0.1$). The displayed twenty digits are those of the exact sector-$0$ cubic \eqref{eq:sector_cubic}, reproducible to arbitrary precision. The operator-level values, from the bordered systems of Appendix~B at $N=24,32,40$ (EP2) and $N=24$--$28$ (EP3) with residuals below $3\times10^{-28}$, reproduce all twenty digits at $N=24$ and working precision $d=30$. Because the collocation is exact on the polynomial sector the residual error is purely arithmetic, so the operator-level accuracy \emph{decreases} with resolution---about $21$, $20$ and $18$ digits in $s_*$ at $N=24,32,40$---tracking the conditioning $\kappa_2$ of the bordered Jacobian rather than any discretisation error; Appendix~B verifies that the linearised forward-error estimate built on $\kappa_2$ reproduces these counts to about one digit, and quantifies the run-to-run fluctuation of the counts themselves. The splitting exponent is measured over $\epsilon\in[10^{-10},10^{-2}]$.}
\label{tab:certified}
\begingroup
\small
\begin{tabularx}{\linewidth}{@{}p{0.30\linewidth}p{0.36\linewidth}X@{}}
\hline
Quantity & Certified value & Cross-check \\
\hline
EP2 ($De=0.15$): $\Lambda_*$ & $1.5095752865775080251$ & operator: all 20 digits \\
EP2 ($De=0.15$): $y_*=|\omega_*|/k$ & $1.2385130101588585499$ & operator: all 20 digits \\
EP2 splitting exponent & $0.50000$ & conjugate-pair symmetric \\
Cusp: $De_*$ & $0.20329171841712946742$ & Eq.~\eqref{eq:deborah_migration}: residual $0$ to 40 digits at working precision (the 20 tabulated digits reproduce it to $2\times10^{-20}$) \\
Cusp: $\Lambda_*$ & $1.3703990989252106724$ & two independent seeds \\
Cusp: $y_*$ & $1.7063465116142830416$ & \\
\hline
\end{tabularx}
\endgroup
\end{table}

\begin{figure}[htbp]
\centering
\includegraphics[width=0.74\textwidth]{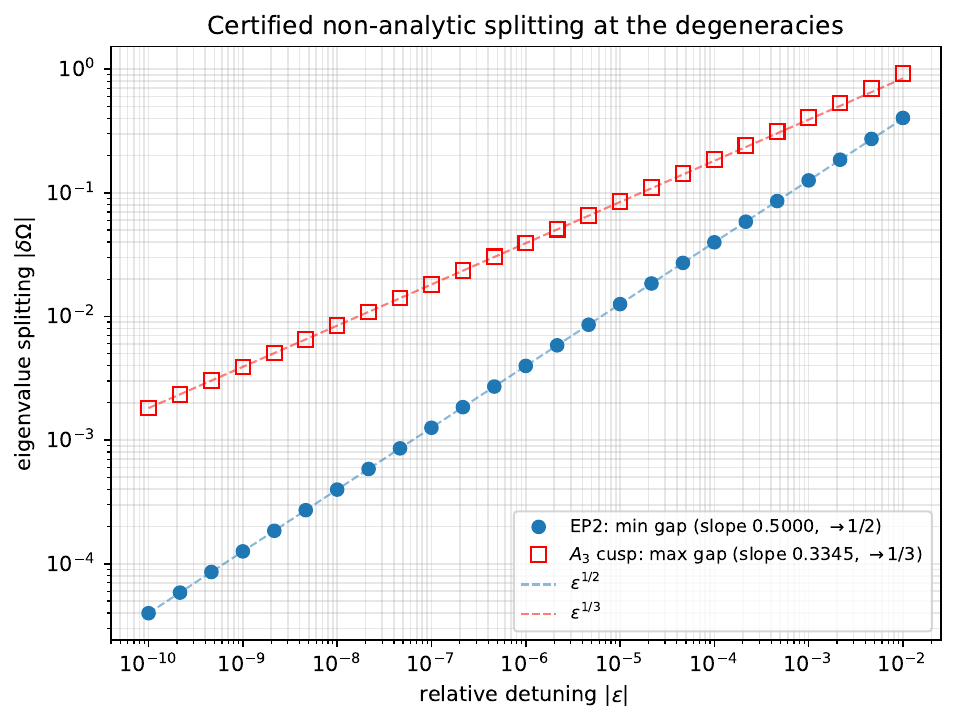}
\caption{Certified non-analytic splitting at the degeneracies of the open system. The two series are different functionals of the spectrum, as the archived figure legend records: the blue series is the \emph{minimum} gap between pairs of roots---the only one that scales as $\epsilon^{1/2}$ at an EP2, the maximum gap being $\mathcal{O}(1)$---and the red series is the \emph{maximum} gap among the three roots at the cusp. Both are plotted on the common axis $|\delta\Omega|$, which denotes the gap in question in each case. Blue circles: eigenvalue splitting under relative detuning $\epsilon$ of $\zeta_0$ off the certified EP2 ($De=0.15$), evaluated on the exact sector-$0$ cubic \eqref{eq:sector_cubic} in multiprecision; by the Theorem of section~\ref{sec:sector_theorem} this carries the same spectrum as the operator pencil, and the operator-level calculation at $N=32$ reproduces the same exponent (block~[D] of \texttt{hyperboloidal\_pencil.py} in the repository, which performs the corresponding sweep---taken there with the opposite sign of $\epsilon$---by multiprecision Newton iteration on the discretised pencil rather than on the sector cubic). The exponent fitted over the whole asymptotic range $\epsilon\le10^{-4}$ is $0.50000$; the locally fitted slope attains that value to five decimals for $\epsilon\le10^{-5}$, and the sweep is taken with $\epsilon>0$, for which the pair remains on the imaginary $\Omega$ axis as two distinct purely imaginary frequencies (for $\epsilon<0$ it leaves the axis as $\Omega_2=-\Omega_1^{*}$; the exponent is the same). Open red squares: splitting of the three roots of the exact sector-$0$ cubic \eqref{eq:sector_cubic} under detuning of $\zeta_0$ off the $A_3$ cusp, following the $\epsilon^{1/3}$ law (guide lines as indicated). The fitted cusp slope is $0.3345$ rather than $0.3333\ldots$; this is a finite-interval fit, not a discrepancy. The exponents $1/2$ and $1/3$ are established algebraically by the Puiseux structure of a double and of a triple root, and the numerical series is evidence of compatibility with them rather than a five-decimal certification of the exponent itself. Both series are algebraic evaluations in $30$-digit arithmetic and therefore carry no discretisation error; double-precision computations of the same quantities lose convergence near the degeneracies, where eigenvalue condition numbers reach $\sim10^{14}$.}
\label{fig:certified_scaling}
\end{figure}

Beyond static certification, the defective character is verified dynamically (figure~\ref{fig:certified_scaling}): perturbing $\Lambda$ off the EP2 by relative amounts $\epsilon$ spanning eight decades yields eigenvalue splittings $|\delta\Omega| = C\,\epsilon^{1/2}$ with locally fitted exponent $0.50000$ throughout the asymptotic range $\epsilon\le10^{-5}$, rising to $0.5096$ only at the largest detuning sampled, where the subleading correction is no longer negligible, and the character of the split pair depends on the sign of the detuning, and we state which is which. Increasing $\Lambda$ off the EP2 ($\epsilon>0$) moves the system into the wedge $\Delta<0$, where all three roots in $y$ are real and the pair remains on the imaginary $\Omega$ axis as two distinct purely imaginary frequencies; decreasing it ($\epsilon<0$) sends the pair off the axis as $\Omega_2=-\Omega_1^{*}$, symmetric about the imaginary axis---equivalently, the $y=i\Omega$ roots form a complex-conjugate pair, as required by the reality of the coefficients in \eqref{eq:open_real_form}. The exponent $1/2$ is the same on both sides; only the geometry of the separation differs. The EP3 solver converges to the analytic cusp from well-separated seeds, providing an independent validation of the triple-degeneracy conditions at the operator level.

\subsection{Robustness under a controlled family of transport-profile deformations}
\label{sec:robustness}

\begin{figure}[htbp]
\centering
\includegraphics[width=0.9\textwidth]{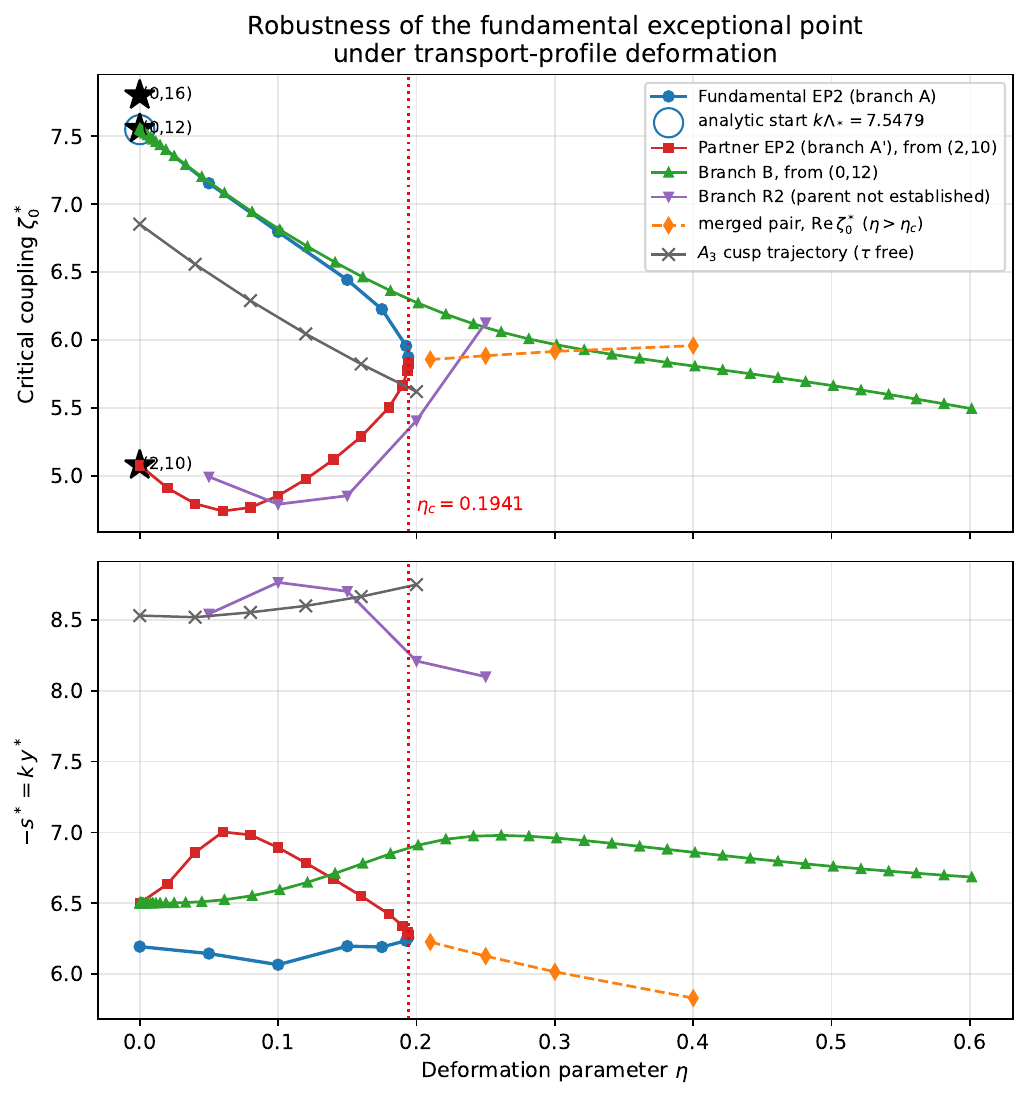}
\caption{Robustness of the fundamental exceptional point of the open system under the transport-profile deformation \eqref{eq:deformation}, at fixed $\tau=0.03$ ($De=0.15$). Top: critical coupling $\zeta_0^*$; bottom: degenerate eigenvalue $-s^* = k y_*$. Branch A is the fundamental (sector-$0$) exceptional point, exact at $\eta=0$ (open circle, at the analytic value $k\Lambda_*=7.5479$); it persists on the imaginary axis at real coupling throughout $\eta\in[0,\eta_c)$ and terminates at $\eta_c=0.1941(1)$. The exact birthplaces at $\eta=0$ of the defective inter-sector crossings \eqref{eq:diabolic_family} of sector pairs $(2,10)$, $(0,12)$ and $(0,16)$ are marked by stars; branch A$'$ is continued to the $(2,10)$ crossing and branch B to $(0,12)$. Branch R2 is drawn for completeness but is \emph{not} attributed to any of the marked crossings: its archived continuation does not reach $\eta=0$ and its parentage is left open (section~\ref{sec:robustness}(iii)). These branches, together with the complex-conjugate escape of the merged pair for $\eta>\eta_c$ and the $A_3$ cusp trajectory, are followed by the same multiprecision continuation and tabulated in the repository. Data from multiprecision bordered-Newton continuation ($N=20$--$28$, $25$--$30$ digits; endpoints certified against exact sector values).}
\label{fig:robustness}
\end{figure}

The exact solvability above rests on the alignment of the viscosity profile with the potential. The physically decisive question---the one raised by the generic non-factorisability of radiative and viscous dissipation---is what survives when this alignment is broken. We deform along a one-parameter family, and are explicit that this is a controlled sample of misalignment profiles rather than a generic perturbation of the operator: the deformation below preserves the parity symmetry $\sigma\to-\sigma$, satisfies $c(\pm1)=1$ and so leaves the indicial structure at the endpoints untouched, and couples only sectors of equal parity with $\Delta m=\pm2$. An odd term would break parity, a profile not approaching unity at the endpoints could alter the Frobenius problem, and a multi-harmonic profile would activate wider couplings. What follows is therefore a statement about this family, not a theorem about generic perturbations---which is nonetheless considerably more than the fine-tuned case $\eta=0$. We therefore deform
\begin{equation}
\zeta(x) \;=\; \zeta_0\,\mathrm{sech}^2 x\,\big(1 + \eta\,\mathrm{sech}^2 x\big)
\quad\Longleftrightarrow\quad
\mathcal{C} = 1 + \eta\,(1-\sigma^2) ,
\label{eq:deformation}
\end{equation}
which destroys the triangular sector structure at $\mathcal{O}(\eta)$, couples the overtone sectors pairwise ($\Delta m = \pm 2$), and makes the effective radiative rate of each hybridised mode depend on $(De,\Lambda)$: precisely the generic regime in which the single-mode reduction is not exact. Polynomial truncation no longer occurs for $\eta\neq0$, and the epistemic consequence is worth stating: the deformation raises the degree by two, since $\zeta_0s\,\mathcal{C}\sigma^m=\zeta_0s(1+\eta)\sigma^m-\zeta_0s\eta\sigma^{m+2}$, so $\mathcal{P}_N$ is no longer invariant and the collocation is no longer exact. The digits quoted in this section therefore carry discretisation error as well as arithmetic error, and are of a different kind from those of table~\ref{tab:certified}---which is why resolution stability, rather than agreement with a closed form, is the criterion used here. The boundary-value structure, on the other hand, is unchanged: since $c(\pm1)=1$ for the deformation \eqref{eq:deformation}, the leading coefficient still degenerates at $\sigma=\pm1$ and the indicial equation there remains $r(r+s)=0$, so the collocation scheme continues to impose exactly the same analyticity condition that selected the eigenfunctions in section~\ref{sec:sector_theorem}. All statements below rest on multiprecision continuation of the bordered systems ($N=20$--$28$ at $25$--$30$ working digits, the fundamental branch at $N=28$, $d=30$), carrying the converged eigenvector along each branch; the data are tabulated in the repository, and the resulting exceptional-point landscape is summarised in figure~\ref{fig:robustness}.

\emph{(i) Finite robustness window.} The fundamental (sector-$0$) EP2 persists as a genuine defective degeneracy, purely on the imaginary frequency axis and at real physical coupling---an unconstrained output rather than an imposed restriction, since the bordered system is solved entirely in complex arithmetic with $s$ and $\zeta_0$ free to acquire imaginary parts, and $\operatorname{Im}s^*$ and $\operatorname{Im}\zeta_0^*$ come out exactly zero at every archived point below $\eta_c$, departing from zero only beyond it--- throughout $\eta \in [0,\eta_c)$ with
\begin{equation}
\eta_c = 0.1941(1) \qquad (De = 0.15).
\end{equation}
Along the branch $\zeta_0^*$ decreases smoothly, $7.5479\to7.1533\to6.4423$ at $\eta=0,\ 0.05,\ 0.15$. The degenerate eigenvalue does not: near $\eta\simeq0.11$ the branch executes a rapid turn in which $\mathrm{d}s_*/\mathrm{d}\eta$ becomes numerically unbounded while $\zeta_0^*$ remains regular, so continuation in $\eta$ stalls there and the turn must be crossed either with a coarser step or by reparametrising the branch by $\zeta_0$. That the same analytic branch is recovered beyond it was verified by eigenvector overlap: $|\langle\varphi_a,\varphi_b\rangle|/(\|\varphi_a\|\,\|\varphi_b\|)=0.9997$ across the turn ($\eta=0.105\to0.13$), indistinguishable from the $0.9987$--$0.99999$ obtained between neighbouring points on either side.

\emph{(ii) Annihilation and complex escape.} At $\eta_c$ the fundamental EP2 merges with a partner exceptional point of the deformed landscape (branch A$'$)---identified by continuing that branch back to $\eta=0$, where it lands on the exact defective inter-sector crossing of sectors $(m,M)=(2,10)$, $(s,\zeta_0)=(-13/2,\ 6601/1300)$, see \emph{(iii)} below, agreeing with it to eight significant digits. The agreement is limited to eight digits, rather than to the working precision, because the bordered EP2 system is singular there. The reason is worth stating precisely, since the crossing \emph{is} an exceptional point in the sense used throughout this paper---algebraically double, geometrically simple, a Jordan block, as section~\ref{sec:robustness}(iii) establishes in exact arithmetic. What it is not is a \emph{regular zero} of the bordered system used to continue the EP2 manifold: at $\eta=0$ the two coalescing eigenvalues belong to different sectors and the augmented Jacobian degenerates, so the birthplace is approached along the branch rather than solved for directly; the identification rests on that approach together with the fold bracketing below, not on a converged residual at $\eta=0$---and the merged pair leaves the real parameter axis as a complex-conjugate pair of exceptional points in complexified $\zeta_0$. Continued from that birthplace, the partner terminates at $\eta=0.19414$ with $\zeta_0^*=5.8316$, against $\eta=0.19403$ and $\zeta_0^*=5.8749$ for the fundamental branch: the two continuations exhaust their step floor on approach to the same fold. We are careful about what that does and does not establish. Neither stop is an upper bound on $\eta_c$---a continuation that halts, halts \emph{below} the fold---and the two stops differ by $1.1\times10^{-4}$, some five times the step floor $2\times10^{-5}$, so at least one of them is limited by conditioning rather than by the fold itself. The only direct upper bound is $\eta=0.21$, where the complex pair is already observed. What locates $\eta_c$ is a fit of the square-root fold law $\zeta_0^*=\zeta_f\pm C\sqrt{\eta_c-\eta}$ to the archived endpoints of the two branches, which gives
\begin{equation}
\zeta_f = 5.8440, \qquad C = 2.754, \qquad \eta_c = 0.19416 ,
\label{eq:fold_fit}
\end{equation}
with residual $6.8\times10^{-4}$ at the endpoint not used in the fit, and $\zeta_f$ consistent with $\operatorname{Re}\zeta_0^*(\eta=0.21)=5.8561$ read off the complex branch. This is how $\eta_c=0.1941(1)$ is obtained. For example, at $\eta=0.21$: $\zeta_0^* = 5.8561 + 0.3365\,i$; at $\eta=0.40$: $\zeta_0^* = 5.9582 + 0.8468\,i$. The transition is a square-root branch point of the EP \emph{location} in parameter space, not an EP3: the coalescing pair separates as $\sqrt{\eta_c-\eta}$ while the third root remains at finite distance, and the EP3 solver fails to converge from seeds on either side. Physically, for $\eta > \eta_c$ the sharp critical-damping transition of the fundamental pair is replaced by a smooth crossover whose sharpness is controlled by $\mathrm{Im}\,\zeta_0^*(\eta)$: the imprint of the nearby complex exceptional point.

\emph{(iii) Creation from defective inter-sector crossings.} The deformation also \emph{creates} exceptional points that the single-mode normal form cannot see. At $\eta=0$, any two sectors $m$ and $M$ of equal parity share an eigenvalue at
\begin{equation}
s \;=\; -\tfrac{1}{2}(m+M+1),
\qquad
\zeta_0 \;=\; \frac{(1+s\tau)\big(s^2+(2m+1)s+m(m+1)+V_0\big)}{-s}\, ,
\label{eq:diabolic_family}
\end{equation}
an algebraically double but geometrically simple (defective) inter-sector crossing.

These crossings and the resonant couplings \eqref{eq:resonant_coupling} of section~\ref{sec:sector_theorem} are not two phenomena but one, and stating so closes the classification of all degeneracies at $\eta=0$. Evaluating \eqref{eq:diabolic_family} at $s=-n$ returns \eqref{eq:resonant_coupling} identically. The crossing frequency $s=-\tfrac12(m+M+1)$ is a negative integer precisely when $m+M$ is odd, i.e.\ when the two sectors have \emph{opposite} parity, and then $n=\tfrac12(m+M+1)$ and $M=2n-1-m$ is exactly the companion index of the pairing in the proof of the Proposition. The classification follows:
\begin{itemize}
\item \emph{equal parity} ($m+M$ even): $s$ is a half-integer, never a resonance; the recurrence closes at the lower index and yields one eigenfunction---\textbf{defective}, a Jordan block. These are the crossings exploited in this section.
\item \emph{opposite parity} ($m+M$ odd): $s=-n$ is a resonance, with companion $m'=2n-1-m$; two polynomial eigenfunctions of opposite parity span the solution space---\textbf{diabolic}, semisimple. These are precisely the ``harmless'' couplings $m\le2n-1$ of Lemma~3.
\item $m\ge2n$: the companion index is negative, no partner sector exists, and the only polynomial is $(1-\sigma^2)^nq_{m-2n}$, which realises the incoming exponent and is not a quasinormal mode.
\end{itemize}
The resonant set that Lemma~3 must clear out of the way is therefore the \emph{diabolic} half of the inter-sector crossing family, and the set exploited here is the \emph{defective} half. That is also why Lemma~3 is needed exactly where it is: the degeneracies that obstruct the spectral identity are the semisimple ones, and the degeneracies that generate new exceptional points under deformation are the defective ones. Algebraic multiplicity is well defined here because, by Lemma 1, $T(s)$ leaves the finite-dimensional space $\mathrm{span}\{1,\sigma,\dots,\sigma^N\}$ invariant for every $N$, and is triangular on it with the $P_j(s)$ on the diagonal; for any $N\ge M$ the characteristic determinant of that restriction is $\prod_{j\le N}P_j(s)$, which acquires a double root when $P_m$ and $P_M$ vanish together. (We use this only to count algebraic multiplicity on an invariant subspace of finite dimension, not to characterise the spectrum, which is established by the Proposition.) That the crossing is nevertheless geometrically simple follows directly from the recurrence \eqref{eq:recurrence}: when $P_m(s)=P_M(s)=0$ with $m<M$ of the same parity, the Lemma applies at the \emph{smallest} such index, so the series closes at degree $m$ and yields a single polynomial eigenfunction; the geometric multiplicity is therefore one and the degeneracy is a Jordan block rather than a semisimple (diabolic) one. This is not left as an inference from the recurrence: the null space of $T(s_*)$ at each of the three crossings used below is computed in exact rational arithmetic, at truncation orders $M=20,30,40,60$, and has dimension exactly one in every case; the same conclusion is reached independently on the Chebyshev-discretised operator, where the second-smallest singular value stands $10^6$--$10^7$ above the double-precision floor. Both computations are archived. The distinction matters quantitatively and not only in name: a semisimple crossing would open \emph{linearly} in $\eta$, whereas the branches born from these points are genuine EP2s with square-root splitting---which is what the continuation below observes. For $\eta>0$ the parity-preserving inter-sector coupling ($\Delta m = \pm 2$) resolves these defective crossings into genuine EP2 branches---consistent with their defective character, since a semisimple crossing would instead open linearly in $\eta$. Two members of this family were followed by multiprecision continuation, together with a third branch whose origin is left open (all three tabulated in the repository): the $(0,12)$ branch born at $(s,\zeta_0)=(-6.5,\ 7.5546\ldots)$ (branch B, followed to $\eta=0.60$); a further deformed sectoral branch followed over $\eta\in[0.05,0.25]$, whose parentage we leave open: the archived continuation does not reach $\eta=0$ and $\zeta_0^*$ along it is not monotone, so neither the $(0,16)$ crossing at $(-8.5,\ 7.8006\ldots)$ nor the sector-$5$ exceptional point at $(-8.0872,\ 2.9682)$ can be confirmed as its $\eta\to0$ limit; and the $(2,10)$ branch born at $(-6.5,\ 5.0777\ldots)$, which is precisely the annihilation partner of the fundamental exceptional point in \emph{(ii)}. Creation and annihilation in the deformed landscape are thus organised by a single, exactly classified family of defective inter-sector crossings.

\emph{(iv) Persistence of the cusp.} The $A_3$ organizing center itself is robust: continued in $\eta$ with $\tau$ free, the triple degeneracy persists at least to $\eta = 0.20$---the extent of the archived run, with $De_*(\eta)$ varying by about one percent ($\tau_* \in [0.0407,\,0.0411]$) while $\Lambda_*(\eta)$ decreases smoothly. We quote $0.20$ rather than the larger value the continuation code reaches ($\eta\simeq0.44$, with $\tau_*$ still in $[0.0406,\,0.0411]$) because only the former is archived at the certification standard of Appendix~B; the cusp is therefore at least as robust as the fundamental EP2, whose window closes at $\eta_c=0.1941(1)$.

The combined picture is a quantitative answer to the robustness question: the discriminant skeleton of the open normal form is not solely a consequence of exact solvability---it persists throughout the tested deformation class over a finite, explicitly computable window---and the mechanisms by which it reorganises beyond that window (annihilation of exceptional points into complexified parameter space; creation from defective inter-sector crossings) are themselves naturally described in the same singularity-theoretic language. We put it that way deliberately: exhibiting a fold in the exceptional-point location and EP2 branches emerging from defective crossings is evidence that such a description is apt, but we have not constructed the deformation space and classified its singularities, which is what an organising-centre claim at that level would require.


\section{Dynamical interpretation of discriminant crossing}

The global discriminant analysis shows that the parameter plane $(De,\Lambda)$ is partitioned by the curve $\Delta(De,\Lambda)=0$ into regions with distinct algebraic multiplicity structure, yielding a direct dynamical interpretation in the context of relativistic stellar oscillations.

\subsection{Mode classification away from degeneracy}

In the non-dissipative limit where both causal relaxation and viscous dissipation vanish ($De \to 0, \Lambda \to 0$), the system reduces to a quadratic self-adjoint problem describing propagating gravitational–fluid oscillations with purely real stable frequencies \cite{Chandrasekhar1964, LindblomDetweiler1983}. However, for a finite viscous coupling in the purely parabolic Navier--Stokes limit ($De \to 0, \Lambda \neq 0$), the reduced operator becomes a quadratic non-self-adjoint dissipative system where the modes are intrinsically complex. For finite relaxation time ($De>0$), the full cubic spectral surface comprises three sheets:

\begin{itemize}
\item[(i)] Two sheets continuously connected to the non-dissipative oscillatory modes,
\item[(ii)] One sheet associated with the non-hydrodynamic relaxation channel \cite{HiscockLindblom1983}.
\end{itemize}

In this regime the operator remains diagonalisable and the dynamical classification of modes is stable under small parameter variations.

\subsection{Crossing the discriminant curve}

Crossing the curve $\Delta=0$ corresponds to the coalescence of two eigenvalues and the formation of an exceptional point \cite{Heiss2004, Rotter2009}. At such points:

\begin{itemize}
\item The algebraic multiplicity exceeds the geometric multiplicity,
\item The polynomial pencil, restricted to the relevant invariant sector, becomes defective---we assert this for the sector cubic and its companion linearisation, and for the discretised operator, not for a closed infinite-dimensional realisation, whose domain and norm this work deliberately does not fix,
\item The spectral projection onto individual modes ceases to be well defined.
\end{itemize}

Dynamically, this implies that two previously distinct modes become inseparable under infinitesimal perturbations of transport parameters.

In particular, when one of the merging branches is continuously connected to the relaxation mode, the crossing of the discriminant curve signals a reorganisation between:

\begin{itemize}
\item A propagating gravitational–fluid oscillation,
\item A relaxation-dominated non-hydrodynamic mode.
\end{itemize}

The classification into “oscillatory” and “relaxational” branches therefore ceases to be invariant across the EP curve.

\subsection{Cusp and change of branching topology}

At the cusp, where $\Delta=\nabla\Delta=0$, three sheets meet tangentially and the local unfolding acquires cubic-root structure \cite{Arnold1986}: the eigenvalue splitting under transverse perturbations changes from $|\epsilon|^{1/2}$ at generic exceptional points to $|\epsilon|^{1/3}$ there. The spectral sensitivity to transport parameters is correspondingly enhanced, and the local mode structure reorganises according to the $A_3$ geometry; both laws are quantified and verified numerically in section~\ref{sec:sensitivity}.

\section{Spectral sensitivity and structural implications}
\label{sec:sensitivity}

The existence of exceptional points and of a higher-order cusp singularity has direct structural consequences for the spectral problem \cite{Heiss2004, Arnold1986}. These consequences follow from the non-analytic dependence of eigenvalues on control parameters and are independent of detailed microphysical modelling \cite{Kato1966}.

\subsection{Non-analytic spectral response}

In analytic spectral problems, eigenvalues depend smoothly on system parameters \cite{Kato1966}. In contrast, near non-Hermitian degeneracies the dependence becomes non-analytic \cite{Heiss2004, Rotter2009}.

As shown in the previous section, near a generic exceptional point the eigenvalue splitting satisfies
\begin{equation}
\delta \sim |\epsilon|^{1/2},
\end{equation}
whereas in the vicinity of the cusp it obeys
\begin{equation}
\delta \sim |\epsilon|^{1/3}.
\end{equation}

The eigenvalues therefore fail to depend analytically on the control parameters $(De,\Lambda)$ and instead exhibit fractional-power branching \cite{Arnold1986}. This behaviour reflects the local topology of the spectral surface and is intrinsic to the cubic non-Hermitian structure \cite{Heiss2004}. Both scaling laws are verified numerically over eight decades of detuning in figure~\ref{fig:certified_scaling}: both on the exact sector cubic \eqref{eq:sector_cubic}, which by the Theorem of section~\ref{sec:sector_theorem} carries the spectrum of the operator pencil \eqref{eq:pencil}, with the square-root exponent independently reproduced at operator level by multiprecision Newton iteration at $N=32$.

\subsection{Reorganisation of spectral sheets}

Away from degeneracies the three sheets admit a continuous labelling inherited from the non-dissipative limit. Along the exceptional-point curve two of them coalesce and exchange analytic continuation, the geometric multiplicity dropping below the algebraic one; at the cusp all three meet tangentially and the local Riemann surface acquires cubic-root topology \cite{Arnold1986, Heiss2004}.

\subsection{Structural consequence of causal transport}

The emergence of this singular geometry is not model-specific. It follows from a structural property:

\begin{quote}
A second-order dynamical system coupled to a single first-order relaxation channel generically produces a spectral problem of cubic \emph{grade}.

\noindent
The distinction is worth stating in one line, because the two halves of this paper rest on different halves of it: \emph{temporal order fixes the grade of the pencil; the coefficient structure fixes the exceptional-point and cusp geometry}. The counterexample below shows that the first does not imply the second.
\end{quote}

It is important not to overstate what the degree count delivers. Temporal order fixes the \emph{grade} of the pencil; it does not fix its discriminant, and a cubic family need not possess any real-parameter degeneracy at all. The one-parameter family $\lambda^3+\lambda+1+a^2+b^2$, for instance, is cubic in $\lambda$ with two real controls, yet its discriminant $-27(1+a^2+b^2)^2-4$ is strictly negative for every $(a,b)\in\mathbb{R}^2$: no multiple root, no exceptional-point curve, no cusp. What produces the EP2 locus and the $A_3$ organizing center is the specific coefficient structure that a single scalar relaxation channel imposes---the family \eqref{eq:open_polynomial}---and not the temporal order by itself. Within \emph{that} family the exceptional-point curve and its terminating cusp are unavoidable features of the parameter plane \cite{IsraelStewart1979, HiscockLindblom1983}, and section~\ref{sec:exact_realization} exhibits an open boundary-value problem that realises the family exactly.

In the limit $De \to 0$, corresponding to vanishing relaxation time, the cubic structure collapses into a quadratic non-self-adjoint dissipative problem \cite{LandauLifshitzFluid}. Within this parabolic framework, while the higher-codimension $A_3$ cusp singularity is dissolved as the relaxation channel closes, secondary non-Hermitian degeneracies persist, retaining a regular second-order exceptional point (EP2) at $\Lambda = 2$ where the roots coalesce into a double defective eigenvalue $\Omega = -i$. Finite causal relaxation therefore modifies not merely the damping rates but shifts the higher-order topological classification of the non-Hermitian spectral problem.

\subsection{Implications for relativistic stellar dynamics}

If a physical system explores regions of parameter space near the exceptional-point curve, the fractional-power branching implies enhanced spectral sensitivity to variations in bulk transport coefficients \cite{AlfordMahmoodifarSchwenzer2012}. To verify whether realistic stellar configurations dynamically access such non-Hermitian degeneracies, the dimensionless normal form must be mapped to physical transport scales.

\subsection{Mapping to post-merger transport scales}
\label{sec:transport_scales}

A change of object takes place at this point and should be stated plainly. The exactly solvable system of section~\ref{sec:exact_realization} is a $(1{+}1)$-dimensional P\"oschl--Teller model whose radiative rates are fixed by the potential depth, $\gamma_m=(m+\tfrac12)/k$; its role was to establish that the open cubic normal form, its EP2 curves, its cusp and the migration law \eqref{eq:deborah_migration} are realized by genuine Siegert eigenvalues rather than postulated, and to fix the robustness window under profile deformation. What follows is not a statement about that model. The criterion derived below is a statement about the EP2 locus of the open normal form \eqref{eq:open_polynomial} itself, evaluated at a radiative rate $\gamma_{\rm GW}$ appropriate to gravitational-wave emission from a stellar remnant and with the mode speed identified as $c_s=v_\beta$; it inherits from section~\ref{sec:exact_realization} the guarantee that the locus is that of a genuine radiating problem, but not the specific numerical values of $V_0$ or $k$ used there. The fiducial radiative rate is, however, realizable within that same family: $\gamma_0=1/(2k)=0.03$ requires $k=50/3$ and hence $V_0=k^2+\tfrac14=278.03$, a perfectly admissible potential depth. The value used below is therefore not merely illustrative in the abstract, but attainable by a member of the exactly solvable class constructed above.

To determine whether the migrated open-system geometric structure derived in section~\ref{sec:organizing_center} is accessible to realistic compact objects, we evaluate the generalized polynomial Eq.~\eqref{eq:open_polynomial} using parameters representative of a hypermassive neutron star remnant undergoing highly compressive core oscillations, such as core composition $g$-modes or radial compression modes. The landmark observation of binary neutron star mergers via gravitational waves \cite{Abbott2017} and the advent of third-generation detectors \cite{Maggiore2020} have motivated intensive numerical relativity efforts to model viscous transport in merging remnants \cite{Radice2022, Ripley2023}. State-of-the-art nonlinear simulations indicate that out-of-equilibrium weak interactions and out-of-equilibrium Urca processes can generate substantial bulk viscosities, effectively damping the remnant's violent density oscillations \cite{Most2024, Camelio2024, ChabanovRezzolla2025, Alford2025}.

The model considers a post-merger configuration with mass $M = 2.7 M_\odot$ and a thermally inflated radius $R = 15$ km, oscillating in its fundamental compressive mode at $f_0 \approx 3.0$ kHz \cite{Radice2022}, within the range of peak frequencies reported in numerical-relativity surveys. We fix the radiation leakage parameter driven by gravitational wave emission at $\gamma_{GW} \approx 0.03$, corresponding to an amplitude damping time $\tau_{GW}=1/(\gamma\omega_0)\simeq1.8$~ms at $f_0=3$~kHz. This is an illustrative value rather than one derived for the chosen mode, and we note an internal tension with section~\ref{sec:nogo}, where the compressive modes to which bulk viscosity couples are characterised by $\gamma\lesssim10^{-2}$. The two are not reconciled here. The choice is, however, conservative in the direction that matters: $\Theta(\gamma)$ \emph{increases} as $\gamma$ decreases, towards $\Theta(0)=8$, so a smaller and more representative $\gamma$ would make the exceptional point harder to reach and would strengthen, not weaken, the negative conclusion of section~\ref{sec:nogo}.

Two physical requirements must be imposed before any parameter is chosen, and together they nearly determine the fiducial point. The first is a identification of the mode wavenumber. For a fundamental compressive mode with a node at the stellar surface we take $k_{\rm eff}=\pi/R$ rather than $1/R$. This is a modelling choice---the fundamental wavenumber of a homogeneous acoustic cavity with that node structure---and not a quantity derived from a relativistic eigenmode: a stellar mode has a stratified $c_s(r)$, curvature, a non-sinusoidal eigenfunction and surface conditions of its own, and a composition $g$-mode does not even obey an acoustic dispersion $\omega=c_sk$. We adopt it because it is the least unreasonable single number available, and because the alternative $1/R$ is excluded outright by causality; the corresponding phase speed is $c_s=\omega_0/k_{\rm eff}=2f_0R$, which for the parameters above gives $c_s\simeq0.300\,c$. (The cruder identification $k_{\rm eff}\sim1/R$ would give $c_s\simeq0.94\,c$ and is untenable.) The second is causality of the underlying Israel--Stewart theory itself: the characteristic speed of the bulk channel, $v_\Pi^2=\zeta/(\rho\tau)$, with $\rho$ the mean density (formed from the gravitational mass; this is a fiducial inertia, not the baryonic rest-mass density, and the criterion below is insensitive to the distinction) $\bar\rho=3M/(4\pi R^3)=3.80\times10^{14}$ g cm$^{-3}$, must satisfy
\begin{equation}
v_\Pi^2 + c_s^2 \;\le\; c^2 .
\label{eq:IS_causality}
\end{equation}

The choice of inertia is immaterial here: had the relativistic inertia $(\varepsilon+p)/c^2$ been used in place of $\rho$, both $\chi$ and $\rho c_s^2$ would rescale by the same factor, so \eqref{eq:ratio_kinematic} and \eqref{eq:nogo_sound} below are invariant under that choice.

These two requirements interact with the geometry of section~\ref{sec:cusp} in a way that is not merely restrictive but predictive, because the dimensionless ratio of the two control parameters is itself a ratio of characteristic speeds,
\begin{equation}
\frac{\Lambda}{De} \;=\; \frac{\zeta}{\rho\,\tau\,c_s^2} \;=\; \frac{v_\Pi^2}{c_s^2} .
\label{eq:Lambda_over_De}
\end{equation}
Along the EP2 locus this ratio is bounded below: in the closed limit $\Lambda/De=(y^2+1)^2/(y^2-1)$, minimised at $y=\sqrt3$ with value exactly $8$, and at $\gamma_{GW}=0.03$ the minimum is $7.596$, attained at $y_*=1.7227$. Combining this with \eqref{eq:IS_causality} and \eqref{eq:Lambda_over_De} yields a sharp necessary condition, independent of the equation of state:
\begin{equation}
c_s \;\le\; \frac{c}{\sqrt{1+\min(\Lambda/De)}} \;\simeq\; \frac{c}{3} .
\label{eq:cs_bound}
\end{equation}
\emph{A causally accessible exceptional point requires a mode speed below approximately $c/3$}, and the least demanding configuration is precisely the $A_3$ cusp. That coincidence is not numerical: it is forced by the geometry, and we record the reason, since it is what converts the cusp from a classificatory curiosity into the operative object of this section.

\medskip
\noindent
\textbf{Lemma (the cusp is the point of minimum transport cost).}
\textit{Along the EP2 locus \eqref{eq:EP_param_gamma} at fixed $\gamma$, the three functions $De(y)$, $\Lambda(y)$ and the control ratio $R(y)\equiv\Lambda/De$ are stationary at the same point, and that point is the $A_3$ cusp. It is a strict minimum of $R$.}

\noindent\textit{Proof.}
Differentiating \eqref{eq:EP_param_gamma} and clearing denominators,
\begin{equation}
\begin{aligned}
\mathrm{num}\,De'(y) &= -G(y;\gamma),\\
\mathrm{num}\,\Lambda'(y) &= \big[(y-\gamma)^2+1\big]\,G(y;\gamma),\\
\mathrm{num}\,R'(y) &= 2\big[(y-\gamma)^2+1\big]\,G(y;\gamma),
\end{aligned}
\label{eq:cusp_stationarity}
\end{equation}
with the single common factor
\begin{equation}
G(y;\gamma) \;=\; y^3 - 3(1+\gamma^2)\,y + 2\gamma(1+\gamma^2).
\label{eq:cusp_cubic}
\end{equation}
Since $(y-\gamma)^2+1>0$ for every real $y$, the three derivatives vanish together, and only where $G=0$. That $G=0$ is exactly the cusp condition follows by eliminating $De_*$ between $y_*=(1+2De_*\gamma)/(3De_*)$, i.e.\ $De_*=1/(3y_*-2\gamma)$, and the migration law \eqref{eq:deborah_migration}: substituting gives $27\,G(y_*;\gamma)/(3y_*-2\gamma)^3=0$. Hence \eqref{eq:cusp_cubic} is the migration law written in the variable $y$, and at $\gamma=0$ it reduces to $y^3-3y=0$, $y_*=\sqrt3$. Finally $R''(y_*)>0$ on the primary branch---$12.0$, $11.79$, $11.30$, $9.88$, $6.00$ at $\gamma=0,\,0.03,\,0.1,\,0.3,\,1$---and $R'$ changes sign from negative to positive across $y_*$, so the stationary point is a strict minimum. It is moreover the \emph{only} one: $G'(y)=3\big[y^2-(1+\gamma^2)\big]$, which is strictly positive on the admissible branch $y^2>1+\gamma^2$, so $G$ is strictly increasing there and its root is unique. The cusp is therefore the unique stationary point of the control ratio on the admissible branch, not merely a local minimum among several. $\blacksquare$

\noindent
The content of the Lemma is that the $A_3$ cusp is the \emph{singular point of the EP2 plane curve}: $De'(y_*)=\Lambda'(y_*)=0$, so the curve has a genuine cusp there in the sense of plane-curve geometry, not merely a singular point of the discriminant. The consequence is stronger than a homogeneity argument would give. For \emph{any} differentiable $F(De,\Lambda)$ whatsoever,
\begin{equation}
\frac{dF}{dy}\bigg|_{y_*} \;=\; F_{De}\,De'(y_*) \;+\; F_{\Lambda}\,\Lambda'(y_*) \;=\; 0 ,
\label{eq:cusp_universal_stationarity}
\end{equation}
so every differentiable function of the two transport parameters---not merely those homogeneous of degree zero, and whatever the microphysics that fixes it---is automatically stationary along the locus at the cusp. The control ratio is the physically relevant instance. The cusp is therefore not only the organizing center of the parameter plane: it is, necessarily and for geometric rather than accidental reasons, the point of minimum transport cost at which a degeneracy can be reached at all.

A closed control point is available at $\gamma=1$, where the cusp is exactly rational: $De_*=1/4$, $\Lambda_*=1/2$, $y_*=2$ and $\Theta=2$, as \eqref{eq:cusp_cubic} gives $y^3-6y+4=(y-2)(y^2+2y-2)$.

The fiducial point follows. Equation \eqref{eq:cs_bound} gives $c_s\le0.341\,c$, hence $f_0\le3.41$ kHz at $R=15$ km; our choice $f_0=3.0$ kHz satisfies it. Setting $De=De_*=0.1958$, the cusp value at this $\gamma_{GW}$, fixes the relaxation time at $\tau=De_*/\omega_0\simeq0.0104$ ms, and $\Lambda_*=1.487$ then corresponds to $\zeta\simeq2.4\times10^{30}\ \mathrm{g\,cm^{-1}\,s^{-1}}$, with $v_\Pi=0.827\,c$ and $v_\Pi^2+c_s^2=0.775<1$: the configuration is internally causal.

\subsection{A microphysical no-go for Urca-driven matter}
\label{sec:nogo}

Whether such a configuration is realised is a question about the microphysics, and equation \eqref{eq:Lambda_over_De} admits a reformulation that answers it sharply and in purely thermodynamic terms.

The bulk-viscous response of a relaxing medium takes the Mandelstam--Leontovich form $\zeta=\rho(v_{\rm FR}^2-v_\beta^2)\,\tau$, where $v_{\rm FR}$ is the sound speed at \emph{frozen} composition and $v_\beta$ that of the $\beta$-equilibrated (barotropic) fluid. Here $\rho$ is the effective inertia of the compressive mode, the same quantity that normalises \eqref{eq:Lambda_def} and that was evaluated as the mean density $\bar\rho=3M/(4\pi R^3)$ in section~\ref{sec:transport_scales}; as noted there, in \eqref{eq:ratio_kinematic} and \eqref{eq:nogo_sound} the susceptibility is \emph{defined} as $\chi=\rho(v_{\rm FR}^2-v_\beta^2)$, so $\rho$ cancels identically and those two equations are invariant under the choice. The invariance does \emph{not} extend to \eqref{eq:ratio_microphysical}--\eqref{eq:three_channels} below, where $\chi$ is instead an absolute energy density taken from the transport literature and is not proportional to $\rho$: there the ratio $\chi/(\rho c_s^2)$ does shift with the convention, by a factor of order $1.2$--$1.4$ for a $2.7\,M_\odot$ remnant. That shift is immaterial to the conclusion---the shortfall is a factor $\simeq235$---but it is a rescaling of the numbers, not an invariance, and we state it as such. Equivalently, $\zeta=P(\gamma_\infty-\gamma_0)\tau$ in terms of the frozen and equilibrium adiabatic indices, the form used throughout the dense-matter transport literature \cite{Alford2022Particles, AlfordHaber2021, Alford2024Quark}. It is worth recording that this is not an external import but is forced by the pencil analysed here, which in dimensional form reads $(1-i\omega\tau)(\omega^2-c_s^2k^2)+i\omega\zeta k^2/\rho=0$. In the low-frequency limit $\omega\tau\ll1$ it gives $\omega/k\to c_s$, and in the high-frequency limit $\omega\tau\gg1$ it gives $\omega^2/k^2\to c_s^2+\zeta/(\rho\tau)$. Requiring these to be the $\beta$-equilibrated and frozen sound speeds respectively forces $\zeta=\rho\tau(v_{\rm FR}^2-v_\beta^2)$ \emph{with the same $\tau$ that defines $De$}---which settles, for the reduced single-channel pencil analysed here, that the Mandelstam--Leontovich relaxation time is the same $\tau$ that defines $De$. It does not establish that an arbitrary chemical relaxation spectrum of dense matter reduces to a single Israel--Stewart $\tau$; that is an assumption of the closure, not a consequence of it. The susceptibility is therefore $\chi=\rho(v_{\rm FR}^2-v_\beta^2)$, and with $c_s=v_\beta$ for the slow compressive mode the control ratio becomes purely kinematic, One remark on the composition of these identifications, each declared separately above but whose chaining is not: a composition $g$-mode is assigned an acoustic $k_{\rm eff}$, from which $c_s=0.300\,c$ follows, and that number is then identified with the $\beta$-equilibrated sound speed of the matter. The quantity entering \eqref{eq:ratio_kinematic} is therefore derived from a hypothesis about the modal structure rather than measured as a thermodynamic property; as shown below, the conclusion is insensitive to it, and in fact conservative.
\begin{equation}
\frac{\Lambda}{De} \;=\; \frac{v_{\rm FR}^2-v_\beta^2}{v_\beta^2} \;=\; \left(\frac{v_{\rm FR}}{v_\beta}\right)^{\!2}-1 ,
\label{eq:ratio_kinematic}
\end{equation}
so that the criterion $\Lambda/De\ge\Theta(\gamma)$ reads, for weakly damped compressive modes,
\begin{equation}
v_{\rm FR}\;\ge\;3\,v_\beta .
\label{eq:nogo_sound}
\end{equation}
An exceptional point requires the frozen sound speed to exceed the equilibrium sound speed by a factor of three. The factor is $\sqrt{1+\Theta(\gamma)}$ in general and equals three exactly in the closed limit; at the fiducial radiative rate $\gamma_{\rm GW}=0.03$ it is $\sqrt{1+7.596}=2.93$, and it falls further for strongly radiating modes. Since bulk viscosity couples to compressive motion, for which $\gamma\lesssim10^{-2}$, the weak-radiation value $3$ is the relevant one throughout this section and is quoted as such; it is also the most demanding, so using it does not weaken the negative conclusion.

Two consistency checks follow. First, \eqref{eq:ratio_kinematic} identifies the characteristic speed of the bulk channel as $v_\Pi^2=\zeta/(\rho\tau)=v_{\rm FR}^2-v_\beta^2$, so the linear Israel--Stewart causality condition \eqref{eq:IS_causality} for this single bulk channel is exactly $v_{\rm FR}\le c$: the requirement is that the \emph{frozen} sound speed be subluminal. We claim this equivalence only at the linear level and for the closure used here; the full nonlinear causality constraints of Israel--Stewart theory are richer than their linearisation \cite{Cordeiro2025}. This is precisely the thermodynamic stability--causality condition $0<v_\beta^2<v_{\rm FR}^2<1$ imposed independently by Montefusco, Antonelli and Gulminelli \cite{Montefusco2025}, following Camelio \emph{et al.} \cite{Camelio2023}, who identify $v_{\rm FR}$ as the maximal characteristic speed of a chemically reacting mixture. Second, combining \eqref{eq:nogo_sound} with $v_{\rm FR}\le c$ returns $v_\beta\le c/3$, reproducing \eqref{eq:cs_bound} by an independent route.

The reformulation is decisive because $v_{\rm FR}/v_\beta$ is a property of the equation of state, computable without any transport calculation. One qualification is needed on what ``equation of state'' means here: the ratio is \emph{not} determined by a barotropic relation $p(\varepsilon)$ alone. It requires a thermodynamic description carrying the compositional variables explicitly, together with a specification of which reactions are held frozen and which are equilibrated---$v_{\rm FR}$ is the sound speed at frozen composition, $v_\beta$ that of the $\beta$-equilibrated fluid, and the two coincide for a barotrope. Once that specification is made, the ratio is fixed by the equilibrium thermodynamics and is genuinely independent of the relaxation-rate calculation, which is the property we exploit. So specified, it has been evaluated across the Bayesian-allowed space of nucleonic equations of state. Ref.~\cite{Montefusco2025} computes $f$- and $p_1$-mode frequencies in both the frozen and $\beta$-equilibrated limits for a nuclear metamodel posterior constrained by astrophysical observations, nuclear experiment and chiral effective field theory, and finds the two calculations to agree within five per cent for $p_1$ (and well below one per cent for $f$), the agreement degrading with increasing mass. The relevant diagnostic is $p_1$, whose frequency is set by the sound speed, rather than $f$, which scales with the mean density; the quoted agreement therefore \emph{strongly disfavours} an order-unity, mode-averaged difference between the frozen and equilibrated stiffness across the allowed nuclear parameter space. We are deliberate about the strength of this inference: a $p_1$ frequency is a global functional $\omega_{p_1}=F[c_s(r),\rho(r),g(r),\dots]$ of the stellar profile, so a five-per-cent frequency difference is not a pointwise bound on $c_{s,\rm frozen}(r)/c_{s,\beta}(r)$ and we do not claim it as one (the $p_1$ frequency scales linearly with the sound speed at fixed stellar structure, so a five-per-cent frequency difference corresponds to a comparable difference in the mode-averaged ratio; the mapping is not exact, but the required factor is three, so the conclusion is insensitive to it), against the factor of three demanded by \eqref{eq:nogo_sound}. That calculation is for cold catalysed matter, so its transfer to a hot remnant is indicative rather than exact; the margin, however, is two orders of magnitude. The transport calculations agree: the susceptibility quoted below corresponds to $v_{\rm FR}/v_\beta=1.017$ for Urca-driven $npe\mu$ matter. The analogous figures for 2SC quark matter and hyperonic matter, $1.097$ and $1.187$, come from dividing published susceptibilities by the \emph{nucleonic} fiducial stiffness, and are therefore benchmark comparisons rather than the $v_{\rm FR}/v_\beta$ of those compositions---for the reason set out after \eqref{eq:three_channels}. The shortfall is therefore not an artefact of one composition or one transport calculation; no such factor-of-three separation appears in the nucleonic equation-of-state ensemble examined here, and the exotic benchmarks likewise sit far below threshold when measured against the fiducial stiffness---though their same-equation-of-state $v_{\rm FR}/v_\beta$ remains to be computed, and we do not claim a general statement about dense matter.

 In the regime relevant here the bulk-viscous response takes the relaxation form $\zeta=\chi\,\tau$, where $\chi$ is the out-of-equilibrium susceptibility of the dissipative channel (in the notation of Alford, Harutyunyan and Sedrakian \cite{Alford2022Particles}, $\chi=C_1^2/A_1$ for the electronic Urca channel). Substituting into \eqref{eq:Lambda_over_De}, the relaxation time \emph{cancels}:
\begin{equation}
\frac{\Lambda}{De} \;=\; \frac{\zeta}{\rho\,\tau\,c_s^2} \;=\; \frac{\chi}{\rho\,c_s^2} .
\label{eq:ratio_microphysical}
\end{equation}
The control ratio is therefore no longer independent once the constitutive and modal identifications above are imposed---the Mandelstam--Leontovich form, a single bulk channel, $c_s=v_\beta$, and the identification of $\rho$---but becomes a fixed ratio of the non-equilibrium susceptibility to the adiabatic stiffness of the medium. Absent those identifications, $\Lambda$ and $De$ remain independent parameters of the phenomenological model. Combined with the bound $\Lambda/De\ge8$ derived above, \eqref{eq:ratio_microphysical} becomes a criterion on the matter itself:
\begin{equation}
\boxed{\;\chi \;\ge\; 8\,\rho\,c_s^2\;}
\label{eq:nogo}
\end{equation}
\emph{An exceptional point requires the out-of-equilibrium response of the dissipative channel to exceed the adiabatic stiffness of the medium by nearly an order of magnitude.}

The numerical factor in \eqref{eq:nogo} is not universal: it is the minimum of $\Lambda/De$ over the EP2 locus, which depends on the radiative damping of the mode. Writing $\Theta(\gamma)\equiv\min_y(\Lambda/De)$, the minimum being taken over the \emph{admissible primary branch} of the EP2 locus---the branch continuously connected to the closed-limit cusp, on which $De>0$ and $\Lambda>0$; by the Lemma of section~\ref{sec:transport_scales} that minimum is attained at the cusp, the relevant root of \eqref{eq:cusp_cubic}. (Without that restriction the minimisation is not meaningful: for $\gamma\gtrsim1$ the secondary branch $\gamma De>1$ carries points with $\Lambda<0$ and hence negative ratios, which are thermodynamically inadmissible.) The criterion then reads $\chi\ge\Theta(\gamma)\,\rho c_s^2$, with $\Theta(0)=8$, $\Theta(0.03)=7.596$, $\Theta(0.3)=4.868$, $\Theta(1)=2$, and the exact asymptotic law
\begin{equation}
\Theta(\gamma) \;\longrightarrow\; \frac{8}{3\sqrt{3}\,\gamma}
\qquad (\gamma\gg1),
\label{eq:threshold_law}
\end{equation}
whose constant is precisely the closed-limit cusp coupling $\Lambda_*=8/(3\sqrt3)$ of section~\ref{sec:organizing_center}. Strongly radiating modes are thus easier to bring to a degeneracy, the required susceptibility falling as $1/\gamma$. The gain is however modest for the modes of interest here: bulk viscosity couples to compressive fluid motion, and for such modes the radiative damping is weak, $\gamma\lesssim10^{-2}$, so $\Theta$ differs from $8$ by only a few per cent. Equation \eqref{eq:threshold_law} nonetheless completes the criterion, and shows that the relevant quantity is the product $\chi\gamma$ rather than $\chi$ alone whenever the mode is strongly damped.

Evaluating \eqref{eq:ratio_microphysical} for Urca-driven $npe\mu$ matter settles the case. Taking $\chi=C_1^2/A_1\simeq5\times10^{6}$ MeV$^4$ from the neutrino-transparent, $T\simeq5$ MeV results of Refs.~\cite{Alford2023PRD, Alford2022Particles} (their figure 6a; the value rises to at most $\sim10^{7}$ MeV$^4$ at the highest densities considered), and $\rho c_s^2\simeq3.1\times10^{34}$ erg cm$^{-3}$ for the remnant above, one obtains
\begin{equation}
\left.\frac{\Lambda}{De}\right|_{\rm Urca} \;\simeq\; 0.034 \qquad \text{against the required} \qquad \frac{\Lambda}{De}\ \ge\ 8 ,
\end{equation}
a shortfall of more than two orders of magnitude, and still a factor $\sim120$ at the most favourable susceptibility in that data set. Equivalently, at $\tau\simeq0.01$ ms the available bulk viscosity is $\zeta=\chi\tau\simeq10^{28}\,\mathrm{g\,cm^{-1}\,s^{-1}}$, consistent with the resonant maximum reported in Ref.~\cite{Alford2022Particles} once the $\zeta_{\rm max}\propto\omega^{-1}$ scaling to $3$ kHz is applied, and two orders of magnitude below the value the degeneracy requires. We therefore state the conclusion plainly:

None of the channels examined here reaches the exceptional-point regime, and the obstruction cannot be removed by tuning the relaxation time: $\tau$ cancels identically. Nor can it be removed by varying $\omega\tau$ within the same single-channel local relaxation response at fixed equation of state and mode structure. We do \emph{not} claim that changing the mode or the frequency is inconsequential in general: doing so alters $\gamma$, $k_{\rm eff}$, the effective mode speed, the eigenfunction and the susceptibility $\chi(T,\rho,\omega)$, and $\Theta=\Theta(\gamma)$ depends on the first of these.

Susceptibilities are read off published resonant maxima as follows. For a relaxing channel $\zeta(\omega)=\chi\tau/(1+\omega^2\tau^2)$, so that at fixed $\omega$ the viscosity is maximal at $\tau=1/\omega$ with $\zeta_{\max}=\chi/(2\omega)$; hence $\chi=2\omega\zeta_{\max}$. (The relaxation rate $1/\tau$ is meant here; the symbol $\gamma$ is reserved throughout this paper for the radiative damping rate of \eqref{eq:open_polynomial}, and the two should not be conflated.)

Applying this to the three channels for which resonant maxima have been computed gives a consistent picture. For Urca-driven $npe\mu$ matter, $\chi\simeq1.0\times10^{33}$ erg cm$^{-3}$. For strangeness-changing processes in hyperonic matter, whose maximum lies at $T\simeq4$ keV and reaches $\zeta_{\max}\sim10^{30}\,\mathrm{g\,cm^{-1}\,s^{-1}}$ at 1 kHz \cite{AlfordHaber2021}, one obtains $\chi\simeq1.3\times10^{34}$ erg cm$^{-3}$. For neutrino-transparent two-flavour colour-superconducting (2SC) quark matter, whose maximum lies at $T\simeq5$ MeV with $\zeta_{\max}\sim5\times10^{29}\,\mathrm{g\,cm^{-1}\,s^{-1}}$ at $n_B=4n_0$ and $\omega/2\pi=1$ kHz \cite{Alford2024Quark} (an extended study built on the same framework appears in Ref.~\cite{Alford2025}; the two titles differ only in the words \emph{two-color} and \emph{two-flavor color}, and both are as published---these are distinct papers, not a duplicated bibliography entry), $\chi\simeq6\times10^{33}$ erg cm$^{-3}$. The corresponding values of the control ratio are
\begin{equation}
\left.\frac{\Lambda}{De}\right|_{\rm Urca}\simeq0.034, \qquad
\left.\frac{\Lambda}{De}\right|_{\rm 2SC}\simeq0.20, \qquad
\left.\frac{\Lambda}{De}\right|_{\rm hyperonic}\simeq0.41,
\label{eq:three_channels}
\end{equation}
against a requirement of $\Lambda/De\ge8$. Every channel evaluated falls short, by factors ranging from $\simeq240$ down to $\simeq20$.

A caveat governs the last two of these and we state it before drawing any conclusion from them. The identity $\Lambda/De=\chi/(\rho c_s^2)$ requires $\chi$, $\rho$ and $c_s$ to describe \emph{the same medium in the same thermodynamic state}. That condition is met for the Urca case, where susceptibility and stiffness both refer to nucleonic matter at post-merger densities. It is \emph{not} met for the exotic channels: $\chi_{\rm 2SC}$ is read off a quark-matter calculation at $n_B=4n_0$ and $\chi_{\rm hyperonic}$ off hyperonic matter in a different thermal regime, yet both are divided by the single macroscopic $\rho c_s^2\simeq3.1\times10^{34}\,{\rm erg\,cm^{-3}}$ of the nucleonic fiducial of section~\ref{sec:transport_scales}. The entries \eqref{eq:three_channels} for 2SC and hyperonic matter are therefore \emph{susceptibility comparisons against a common nucleonic stiffness}, not the control ratios those equations of state would actually produce, and the shortfall factors $\simeq40$ and $\simeq20$ are shortfalls relative to the fiducial $c_s=0.300\,c$, not no-go theorems for those compositions. The point is not academic: the closing argument of this section proposes precisely a softened phase with $c_s\ll0.3\,c$, and at fixed $\chi$ the ratio grows as $c_s^{-2}$. The clean way to test the criterion on a 2SC or hyperonic equation of state is to evaluate $v_{\rm FR}/v_\beta$---equivalently $\chi/[(\varepsilon+p)c_s^2]$---at the same thermodynamic point, which the published transport calculations do not currently report. The nucleonic Urca no-go, by contrast, is internally consistent and is the one on which we rest the negative conclusion. Two features of \eqref{eq:three_channels} are worth noting. First, the ordering is informative rather than uniform: exotic channels improve on Urca by one order of magnitude, so the shortfall is not an accident of one composition, but neither is it insensitive to the microphysics. Second, the two most favourable cases sit in different environments --- the hyperonic peak requires keV temperatures, which Ref.~\cite{AlfordHaber2021} associates with cold regions or the inspiral of eccentric binaries, whereas the 2SC peak lies at $T\simeq5$ MeV, squarely in the post-merger regime. The most favourable case in the environment where the oscillations are actually observed is therefore quark matter, short of the threshold by a factor $\simeq40$.

Since $\chi$ spans barely one order of magnitude across these very different compositions, while the shortfall is one to two and a half orders, the conclusion is robust against the microphysical uncertainties: none of the three channel families examined here brings dense matter close to \eqref{eq:nogo}. We evaluate three published calculations, not the whole dense-matter transport literature, and state the conclusion at that scope. A factor of a few in $\chi$ is a realistic uncertainty and affects none of these conclusions. Two independent fiducial choices, moreover, both point the conservative way. One is the radiative rate, since $\Theta(\gamma)$ increases towards $\Theta(0)=8$ as $\gamma$ decreases. The other is the mode speed: $c_s=0.300\,c$ follows from the acoustic identification of section~\ref{sec:transport_scales}, whereas a realistic $\beta$-equilibrated sound speed in this density range, $v_\beta\simeq0.4$--$0.6\,c$, would raise $\rho c_s^2$ by a factor $1.8$--$4$ and deepen the shortfall by the same factor. Neither choice was made to favour the negative conclusion, and both do. The remaining freedom is therefore in the denominator of \eqref{eq:ratio_microphysical}, not the numerator.

The value of \eqref{eq:nogo_sound} is that it is a criterion rather than a benchmark: it can be evaluated for any candidate composition from its equation of state alone. As the three-channel comparison shows, the accessible range of $\chi$ is too narrow to close the gap, so the only viable route is a reduction of $\rho c_s^2$, that is, a strong softening of the equation of state. How much softening is required depends on the channel, and the exotic ones are markedly less demanding. At $\Theta(0.03)=7.596$ the criterion is met for $c_s\le0.0201\,c$ with the Urca susceptibility, but for $c_s\le0.0492\,c$ with 2SC quark matter and $c_s\le0.0696\,c$ with hyperonic matter --- suppressions of $c_s^2$ by factors of $224$, $37$ and $19$ respectively, relative to the value assumed above. In the post-merger environment itself the relevant case is the 2SC channel, requiring $c_s\lesssim0.05\,c$; the hyperonic value $0.07\,c$ applies only in the cold regime where that channel peaks. Either is an order of magnitude less severe than the estimate obtained from the Urca channel alone. Such softening is characteristic of a mixed phase or the vicinity of a first-order deconfinement transition, where $c_s^2$ can approach zero, and it is precisely the regime in which critical slowing down enhances relaxation times as well. The exceptional-point criterion \eqref{eq:nogo} is thus most naturally satisfied, if at all, in a remnant probing a softened branch of the equation of state --- a prediction that is quantitative, falsifiable, and specific about where to look.

\begin{figure}[htbp]
\centering
\includegraphics[width=1.0\linewidth]{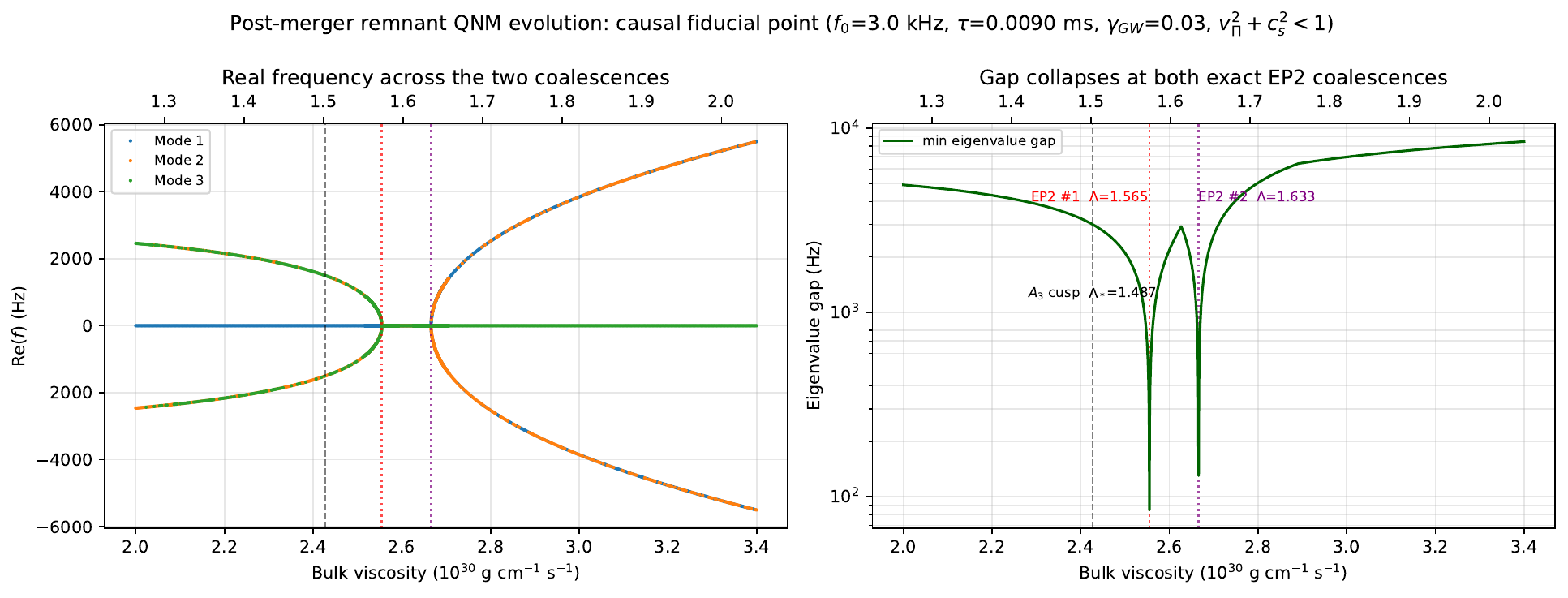}
\caption{Spectral signature of the causal exceptional-point structure, shown for the hypothetical regime in which criterion \eqref{eq:nogo} is satisfied. Parameters follow the causal construction of section~\ref{sec:nogo} ($f_0=3.0$ kHz, $R=15$ km, $\gamma_{GW}=0.03$, $De=0.17$, i.e.\ $\tau=0.0090$ ms, slightly below the cusp value $De_*=0.1958$ ($\tau=0.0104$ ms) quoted in the text, $v_\Pi^2+c_s^2<1$ at both degeneracies). The sweep in $\Lambda$ (upper axes) passes through two real-axis EP2 coalescences at $\Lambda_{\rm EP}=1.565$ and $1.633$, both lying above the $A_3$ cusp coupling $\Lambda_*=1.487$ of the causal fiducial point (dashed line); the panel uses this value of $De$, since at $De_*$ itself the two branches merge and the sweep is tangent to the degeneracy. Left panel: the propagating branches lose their real part as the modes become over-damped through the first coalescence. Right panel: the minimum eigenvalue gap collapses at each defective (Jordan-block) degeneracy. The gap vanishes exactly at the coalescence and cannot be rendered on a logarithmic axis; the dashed vertical lines mark the exact analytic locations, and the plotted curve is the gap evaluated on the sampling grid. As established in section~\ref{sec:nogo}, Urca-driven $npe\mu$ matter falls short of \eqref{eq:nogo} by more than two orders of magnitude; the figure therefore characterises the signature such a configuration would produce, and does not assert that post-merger remnants realise it.}
\label{fig:physical_application}
\end{figure}

Figure~\ref{fig:physical_application} accordingly displays the spectral phenomenology in the hypothetical regime in which \eqref{eq:nogo} is met, at the causal fiducial point constructed above; it should be read as a characterisation of the observable signature that such a configuration would produce, not as a claim that Urca-driven remnants realise it.

The individual modal branches in figure~\ref{fig:physical_application} are tracked across each defective coalescence by robust numerical continuation of the algebraic roots, rescaling the eigenvector at each step to follow the sheets continuously through the Jordan-block point rather than by nearest-neighbour metric assignment, which is ill-defined at a non-Hermitian self-intersection \cite{Rotter2009}. This ensures that the defective coalescences depicted are exact features of the causal algebraic geometry.

\section{Conclusions}

The spectral structure of relativistic stellar oscillations has been analysed in the presence of causal bulk relaxation \cite{IsraelStewart1979, HiscockLindblom1983}. The inclusion of a finite relaxation time promotes the quadratic self-adjoint oscillation problem to a cubic non-Hermitian polynomial eigenvalue problem \cite{Kato1966, Heiss2004}.

Within this framework, the global geometric properties resolve into the following structural hallmarks:

\begin{itemize}
\item The spectrum defines a three-sheeted complex surface over the parameter plane $(De,\Lambda)$ \cite{Arnold1986}.
\item A curve of exceptional points arises from the coalescence of double roots of the cubic polynomial \cite{Heiss2004, Rotter2009}.
\item At a distinguished parameter value, the degeneracy becomes triple and realises an $A_3$ cusp singularity in the sense of Arnold’s classification \cite{Arnold1986}.
\item The local eigenvalue splitting exhibits square-root behaviour near generic exceptional points and cubic-root behaviour at the cusp \cite{Heiss2004}.
\item The open cubic normal form, including its exceptional-point curves, its $A_3$ cusp, and the migration law \eqref{eq:deborah_migration}, is \emph{exactly realized} as the sector decomposition of the quasinormal spectrum of a radiating wave--relaxation system with outgoing Siegert boundary conditions, in which every $\gamma_m$ is an output of the boundary-value problem (section~\ref{sec:exact_realization}).
\item The sector algebra fixes the exceptional points of the open system to twenty significant digits, and multiprecision bordered-Newton solution of the full operator reproduces all twenty; because the collocation is exact on the polynomial sector, the residual discrepancy is purely arithmetic and is purely arithmetic, and the linearised estimate built on the conditioning of the bordered Jacobian tracks it to about one digit. The $0.50000$ square-root splitting exponent is verified over eight decades of detuning on the exact sector cubic, where the fitted value is $0.50000$, and is independently confirmed at five sampled decades---without a five-decimal fit---by multiprecision Newton iteration on the discretised operator.
\item Under the controlled family of transport-profile deformations studied here, which activates inter-mode coupling while preserving parity and the endpoint indicial structure, the fundamental exceptional point persists on the physical domain throughout a finite window $\eta\in[0,\eta_c)$, $\eta_c=0.1941(1)$ at the fiducial parameters, then annihilates with a partner exceptional point of the deformed landscape and escapes into complexified parameter space; the partner and further exceptional points are traced to the exactly classified family of defective inter-sector crossings---with the exception of one further branch, followed but not attributed, whose $\eta\to0$ limit the archived continuation does not reach; and the cusp itself persists with $De_*$ varying by approximately one percent.
\end{itemize}

The logical status of these features deserves a precise statement, since it is easy to claim too much. The coupling of a second-order dynamical system to a single first-order relaxation channel fixes the \emph{grade} of the pencil at three \cite{IsraelStewart1979}; it does not by itself fix the discriminant, and cubic families exist with two real controls whose discriminant never vanishes, hence with no exceptional points at all. What does produce the EP2 curve, the $A_3$ cusp and the migration law is the particular coefficient structure of the family \eqref{eq:open_polynomial} that a scalar relaxation channel imposes, together with the admissibility constraints $De>0$, $\Lambda>0$. Within that family the features above are unavoidable, they are independent of detailed microphysical modelling, and section~\ref{sec:exact_realization} shows the family is not a parametrisation but is realised exactly by an open radiating problem. That is the universality claimed here, and no more.

Two honest delimitations frame these results. First, the exactly solvable system of section~\ref{sec:exact_realization} is a model problem---a scalar wave equation with an effective potential and a minimal relaxation channel---not a perturbed Tolman--Oppenheimer--Volkoff star; the extension to Lindblom--Detweiler polar perturbations with a causal bulk channel is the natural next step, for which the hyperboloidal multiprecision machinery developed here carries over directly. Second, whether realistic dense-matter transport reaches the regime in which these degeneracies occur is a question of microphysics, not of spectral geometry, and section~\ref{sec:nogo} answers it in the negative for the standard case: the criterion $\chi\ge8\rho c_s^2$ is missed by more than two orders of magnitude by Urca-driven $npe\mu$ matter. We regard the criterion itself, rather than that particular verdict, as the transferable result, since it is independent of the relaxation time and can be evaluated for any candidate composition; it locates the required regime unambiguously in strongly softened matter, as near a first-order deconfinement transition. Within these delimitations, the versal unfolding of the $A_3$ organizing center dictates the large-scale arrangement of exceptional-point coalescences in the physical quadrant ($De > 0, \Lambda > 0$), whose observable signature, should the criterion be met, is the one depicted in figure~\ref{fig:physical_application}.

Furthermore, a finite relaxation time does not merely modify damping rates; it fundamentally changes the spectral class of the oscillation problem \cite{Kato1966}. In the limit of vanishing relaxation time ($De \to 0$), the cubic structure collapses into a quadratic non-self-adjoint dissipative Navier--Stokes system where the higher-codimension cusp dissolves but ordinary EP2 points remain embedded \cite{LandauLifshitzFluid, Heiss2004, Rotter2009}. The principal result of this work is thus geometric: causal bulk relaxation induces a non-trivial Riemann-surface structure in the spectral problem of this exactly solvable model, reorganising its non-Hermitian spectral geometry in a way that persists throughout the tested deformation class over the window quantified in section~\ref{sec:robustness}; whether the same mechanism organises the spectrum of realistic self-gravitating configurations is the natural question for future work, consistent with the delimitations stated above.

\appendix
\section{Rigorous Analytic Unfolding of the Open-System $A_3$ Cusp Singularity}
\label{app:cusp_proof}

To formally certify the structural stability of the open-system degeneracy under the Thom--Arnold classification on real manifolds, we must prove that the mapping from the physical transport parameters to Arnold's canonical control parameters constitutes a regular local diffeomorphism mapping physical real coordinates to strictly real canonical parameters.

Let us define the local perturbation coordinates around the migrated critical vector $(De_*, \Lambda_*, \Omega_*)$ as $\mu = De - De_*$ and $\nu = \Lambda - \Lambda_*$. Since $\Omega_*$ sits strictly on the imaginary frequency axis, the spectral perturbation variable $\delta = \Omega - \Omega_*$ undergoes a phase rotation. To isolate a real coordinate for the canonical unfolding, we define the transformed spectral variable $\tilde{\delta}$ via the complex rotation $\delta = i\tilde{\delta}$, with $\tilde{\delta} \in \mathbb{R}$. Performing a multi-variable Taylor expansion on the open polynomial Eq.~\eqref{eq:open_polynomial} while treating $\gamma$ as a fixed background constraint yields the local versal unfolding. Higher-order mixed terms present in the exact identity \eqref{eq:local_expansion}---in particular those in $\mu\tilde\delta^2$ and $\mu\tilde\delta^3$---are omitted here because they do not enter the first jet of the control map and are removable by the smooth transformations described after \eqref{eq:local_expansion}; the omission is therefore a truncation of a controlled expansion, not of an uncontrolled one:
\begin{equation}
P(\tilde{\delta}; \mu, \nu) \approx \frac{1}{6}P_{\Omega\Omega\Omega}^*(i\tilde{\delta})^3 + \left(P_{\Omega De}^*\mu + P_{\Omega\Lambda}^*\nu\right)(i\tilde{\delta}) + \left(P_{De}^*\mu + P_\Lambda^*\nu\right) = 0,
\label{eq:versal_expansion}
\end{equation}
where the direct and mixed partial derivatives evaluated at the open critical stratum are given by:
\begin{align}
P_{De}^* &= -i\Omega_*^3 + 2\gamma\Omega_*^2 + i(1+\gamma^2)\Omega_*, \qquad P_\Lambda^* &= i\Omega_*, \\
P_{\Omega De}^* &= -3i\Omega_*^2 + 4\gamma\Omega_* + i(1+\gamma^2), \qquad P_{\Omega\Lambda}^* = i, \\
\frac{1}{6}P_{\Omega\Omega\Omega}^* &= -iDe_*.
\end{align}

Substituting the derivative evaluations into Eq.~\eqref{eq:versal_expansion} and collecting terms yields:
\begin{equation}
-De_*\tilde{\delta}^3 + i\left(P_{\Omega De}^*\mu + i\nu\right)\tilde{\delta} + \left(P_{De}^*\mu + i\Omega_*\nu\right) = 0.
\label{eq:expanded_versal}
\end{equation}
To reduce Eq.~\eqref{eq:expanded_versal} into Arnold's standard canonical monic form $\tilde{\delta}^3 + u\tilde{\delta} + v = 0$, we divide across by the real negative scaling $-De_*$. This regular algebraic operation defines the coordinate transformation to the canonical control space:
\begin{align}
u(\mu, \nu) &= \frac{i P_{\Omega De}^*}{-De_*}\mu + \frac{-1}{-De_*}\nu = -\frac{i P_{\Omega De}^*}{De_*}\mu + \frac{1}{De_*}\nu, \\
v(\mu, \nu) &= \frac{P_{De}^*}{-De_*}\mu - \frac{-i\Omega_*}{-De_*}\nu = -\frac{P_{De}^*}{De_*}\mu - \frac{i\Omega_*}{De_*}\nu.
\end{align}

To formally establish that this coordinate mapping operates directly on real parameter manifolds, we evaluate the complex components explicitly at the critical stratum. Substituting the purely imaginary critical frequency $\Omega_* = -iy_*$ (where $y_* = \frac{1+2De_*\gamma}{3De_*} \in \mathbb{R}$) into the parameter derivatives yields:
\begin{align}
P_{De}^* &= y_*^3 - 2\gamma y_*^2 + (1+\gamma^2)y_* \in \mathbb{R}, \\
P_{\Omega De}^* &= i\left(3y_*^2 - 4\gamma y_* + 1 + \gamma^2\right) \in i\mathbb{R}.
\end{align}
Differentiating the control variables $(u,v)$ with respect to the physical perturbations $(\mu,\nu)$ determines the explicit components of the linear transformation. Evaluating these relations reveals that the Jacobian coefficients are strictly real-valued quantities over the physical transport domain:
\begin{align}
\partial_\nu v &= -\frac{y_*}{De_*} \in \mathbb{R}.
\end{align}

To identify the individual entries of the Jacobian we record that the numerator of $\partial_\mu u$, given by $f(y_*)$ with $f(y)=3y^2-4\gamma y+1+\gamma^2$, is strictly positive in the physical domain. The non-degeneracy of $J$ itself does not rest on this---it follows from the closed form obtained below---but the sign fixes the orientation of the canonical coordinates. This holds for \emph{every} $\gamma>0$, with no smallness assumption. Indeed $f(\gamma) = 3\gamma^2 - 4\gamma^2 + 1 + \gamma^2 = 1$ identically, and $f'(y) = 6y - 4\gamma > 0$ for $y > 2\gamma/3$. Admissibility of the transport branch, $De_*\gamma<1$, is equivalent to
\begin{equation}
y_* - \gamma \;=\; \frac{1+2De_*\gamma}{3De_*} - \gamma \;=\; \frac{1-De_*\gamma}{3De_*} \;>\; 0 ,
\end{equation}
so $y_*>\gamma>2\gamma/3$ and therefore $f(y_*) > f(\gamma) = 1 > 0$. The Jacobian element cannot vanish anywhere on the admissible branch. We note that this argument is uniform in $\gamma$ and hence covers the entire sector tower $\gamma_m=(m+\tfrac12)/k$ of section~\ref{sec:sector_theorem}, including the overtones with $\gamma_m>\sqrt{3}$.

The transversality of the unfolding requires the real-valued Jacobian matrix $J = \partial(u,v)/\partial(\mu,\nu)$ to be strictly non-degenerate at the critical locus.
\begin{equation}
J = \begin{pmatrix} 
\partial_\mu u & \partial_\nu u \\ 
\partial_\mu v & \partial_\nu v 
\end{pmatrix} = 
\begin{pmatrix} 
-\frac{i P_{\Omega De}^*}{De_*} & \frac{1}{De_*} \\ 
-\frac{P_{De}^*}{De_*} & -\frac{i\Omega_*}{De_*} 
\end{pmatrix}.
\label{eq:correct_jacobian}
\end{equation}
Computing the determinant yields the global topological invariant of the real transformation:
\begin{equation}
\det(J) = \left(-\frac{i P_{\Omega De}^*}{De_*}\right)\left(-\frac{i\Omega_*}{De_*}\right) - \left(\frac{1}{De_*}\right)\left(-\frac{P_{De}^*}{De_*}\right) = -\frac{1}{De_*^2}\left(P_{\Omega De}^*\Omega_* - P_{De}^*\right).
\label{eq:open_jacobian_det}
\end{equation}

Utilizing the exact triple-root relations on the partial components reveals that the internal bracket simplifies to $P_{\Omega De}^*\Omega_* - P_{De}^* = \frac{2\Omega_*^2}{3De_*}(De_*\gamma - 1)$. Substituting this back gives the real determinant:
\begin{equation}
\det(J) = -\frac{2\Omega_*^2}{3De_*^3}(De_*\gamma - 1).
\label{eq:final_real_determinant}
\end{equation}
Since $\Omega_*$ sits strictly on the imaginary axis, its square is strictly negative ($\Omega_*^2 < 0$). In the physically admissible parameter regime, the domain is bounded by $De_*\gamma < 1$, making the scalar factor $(De_*\gamma - 1)$ strictly negative. Consequently, the combined signs guarantee that the determinant is strictly negative ($\det(J) < 0$). This proves that the matrix is strictly non-degenerate, rigorously validating the local mapping via the Inverse Function Theorem on real manifolds.

\section{Multiprecision certification of defective degeneracies}
\label{app:numerics}

This appendix summarises the numerical method used in sections \ref{sec:certification}--\ref{sec:robustness}. The pencil \eqref{eq:pencil} is discretised by Chebyshev collocation on $\sigma\in[-1,1]$ at $N+1$ nodes; regularity at $\sigma=\pm1$ (the outgoing condition in the hyperboloidal frame) requires no boundary rows. Because the operator is strongly non-normal, eigenvalues near degeneracies carry condition numbers up to $\mathcal{O}(10^{14})$ and double-precision spectra do not converge there; all certified quantities are therefore computed in $25$--$30$-digit arithmetic.

An EP2 is located as a regular zero of the bordered system
\begin{equation}
T(s)\varphi = 0,\qquad
T'(s)\varphi + T(s)\varphi_1 = 0,\qquad
c^{\dagger}\varphi = 1,\qquad
c^{\dagger}\varphi_1 = 0,
\label{eq:borderedEP2}
\end{equation}
of dimension $2n{+}2$ in the unknowns $(\varphi,\varphi_1,s,\zeta_0)$, where $\varphi_1$ is the first Jordan chain vector and $c$ a fixed normalisation vector; primes denote $\partial_s$. An EP3 appends $T''\varphi + 2T'\varphi_1 + T\varphi_2 = 0$ and $c^{\dagger}\varphi_2=0$, freeing additionally $\tau$ (dimension $3n{+}3$). Both systems are solved by Newton iteration with exact analytic Jacobians; One word on terminology before the standard itself. We use ``certified'' in the sense of multiprecision numerical validation---small residual at high working precision, quadratic Newton contraction, stability across resolutions, agreement with the closed-form sector value, and a forward-error estimate consistent with the conditioning---and not in the stricter sense of a verified enclosure. We do not compute interval or Krawczyk enclosures, and we do not verify a Newton--Kantorovich hypothesis on a neighbourhood, so no result here carries a machine-checked existence proof. With that reading fixed, we adopt a single certification standard throughout: a converged solution is reported as certified when the residual of the bordered system falls below $10^{-(d-8)}$ at working precision $d$ (i.e.\ $10^{-22}$ at the $d=30$ used throughout, with $10^{-26}$ or smaller actually attained at the tabulated points) and the leading digits are stable across three spatial resolutions (the forward accuracy against the exact sector value is then set by $\kappa_2$, and is quantified in the audit below). This is the criterion used in table~\ref{tab:certified} and in all quoted digit counts. All Jacobian entries are analytic: with $a=1+s\tau$ one has $T'=\tau(s^2-sL_2-L_1)+a(2s-L_2)+\zeta_0\mathcal{C}$, $T''=2\tau(2s-L_2)+2a$ and $T'''=6\tau$, and the derivatives with respect to $\zeta_0$ and (for the EP3 system) $\tau$ are likewise closed-form; no finite differences are used. We note that the third chain condition is written here as $T''\varphi+2T'\varphi_1+T\varphi_2=0$, which is the canonical relation with $\varphi_2$ rescaled by two---admissible because $\varphi_2$ is fixed only by the homogeneous condition $c^\dagger\varphi_2=0$.

Parameter continuation (in $\eta$, $\tau$, or $\zeta_0$) carries the converged eigenvector of the previous step, rescaled to unit maximum entry, as both the Newton seed and the normalisation vector $c$. The step control is as follows: the increment starts at $\Delta\eta=0.05$, is multiplied by $1.4$ after an accepted step (capped at $0.05$) and halved after a rejected one, with the branch terminated when $\Delta\eta$ falls below $2\times10^{-5}$. A step is accepted when Newton converges to the stated tolerance and the predicted point does not jump branch, tested as $|s_{\rm new}-s_{\rm old}|<0.25+3\Delta\eta$ for the EP2 continuation and $|s_{\rm new}-s_{\rm old}|<0.6$ together with $|\tau_{\rm new}-\tau_{\rm old}|<0.01$ for the EP3 (cusp) continuation, the latter having been added after a spurious basin change was detected near $\eta=0.09$. Because the terminus is a square-root fold of the exceptional-point location rather than a point at which Newton simply slows, $\eta_c$ is obtained as the value at which the step floor is reached, bracketed between the last converged point of each of the two branches, and the digits quoted are those stable under variation of the initial step and of $N$. We are explicit that this is a numerical estimate with a quoted uncertainty and not a certified localisation of the fold: certifying it would require solving an augmented system imposing the fold condition directly, or a pseudo-arclength continuation through the turning point, neither of which we carry out. The value $\eta_c=0.1941(1)$ should be read accordingly, and ``certified'' is reserved in this paper for the degeneracies of table~\ref{tab:certified}; the continuation beyond $\eta_c$ is restarted from a complex seed with $\mathrm{Im}\,\zeta_0$ released. Defectiveness is additionally verified dynamically through the splitting exponent measured over eight decades of parameter detuning. \paragraph*{Audit of the certification.} Because a residual criterion is a statement about the
backward error, we report the quantities that convert it into forward digits. At the certified EP2
($De=0.15$), Newton converges in four iterations from a seed perturbed at the seventh digit, with
residuals $2.4\times10^{1}\to1.8\times10^{-7}\to1.1\times10^{-14}\to2.0\times10^{-29}$; the ratio
$\|F_{k+1}\|/\|F_k\|^2$ settles at $0.18$--$0.33$, the quadratic contraction expected at a
\emph{regular} zero and hence direct evidence for the argument of section~\ref{sec:certification}. The
condition number of the bordered Jacobian is
\begin{equation}
\kappa_2(J) = 5.45\times10^{12},\ 3.34\times10^{14},\ 7.67\times10^{15}
\qquad\text{at } N = 24,\,32,\,40,
\end{equation}
growing steeply with resolution, and it is
this factor---not the residual alone---that sets the attainable digit count. Because the
discretisation is exact on the polynomial sector, the discretisation error vanishes identically and
the entire deviation from the exact sector value is arithmetic; the forward accuracy therefore
\emph{degrades with resolution}, tracking $\kappa_2(N)$ rather than improving with it. At $d=30$,
with residuals in the range $2\times10^{-29}$--$3\times10^{-28}$, the bordered solve agrees with the
exact sector value of $s_*$ to about $21$, $20$ and $18$ significant digits at $N=24,32,40$
respectively (and to about $22$, $21$ and $20$ in $\zeta_0^*$), against local linearised forward-error estimates $\kappa_2\|F\|/(\|J\|\,|s_*|)$---the relative normalisation used in the archived script---of $19$--$20$, $18$--$19$ and $17$--$18$ digits. We call these estimates and not a-priori bounds: they come from the linearisation at the computed point, whereas a rigorous bound for a nonlinear system would require control of the Jacobian and its inverse over a neighbourhood, in the manner of Newton--Kantorovich, which we do not establish. The estimate thus tracks the observed accuracy to
about one digit at every resolution, which is the sharpest statement available and is what
identifies $\kappa_2$ as the operative quantity. Agreement between resolutions is about $20$ digits
in $s_*$ for $N=24$ versus $32$ and about $18$ for $32$ versus $40$---in each case equal to the
forward accuracy of the coarser run, as it must be when the error is arithmetic and monotone in
$\kappa_2$. At fixed $N=24$ the accuracy tracks the working precision one-for-one, again as the
error model requires: $16$, $21$, $31$ and $41$ forward digits at $d=25,30,40,50$, with residuals
$\sim10^{-24}$, $10^{-29}$, $10^{-39}$ and $10^{-49}$. The twenty digits quoted in
table~\ref{tab:certified} are therefore reproduced in full by the archived $N=24$, $d=30$ runs and
are not a limit of the method; $d\ge30$ is used throughout because $d=25$ already falls short of
them. All digit counts in this paragraph are quoted to the nearest unit and were measured over five
Newton seeds perturbed at the seventh digit: since the terminal residual sits at the roundoff floor
of the working precision, the counts fluctuate by one to two units between runs and no more
resolution than that should be read into them.

Finally, at $\zeta_0=0$ the exact P\"oschl--Teller frequencies $s=-(m+\tfrac12)\pm ik$ are eigenvalues
of the \emph{discretised} operator to working precision: for $m=0,1,2,3$ at $N=24$ and $d=30$ the smallest
singular value of the real $2n\times2n$ embedding of $T(s)$, evaluated at the exact frequency, lies between $10^{-30}$ and $5\times10^{-29}$ in absolute terms, that is $10^{-33}$--$10^{-32}$ relative to $\|T\|\simeq2\times10^{3}$---the level of the working precision. The individual values are not quoted, since at the roundoff floor they are not reproducible digit by digit. This is the numerical counterpart of the
exactness of the collocation on the polynomial sector, and confirms that the certification measures
arithmetic rather than discretisation error. The auditing script is included in the repository.

All routines and the certified data sets are available in the repository cited below.

\section*{Data availability}
All analytic derivations, numerical routines, certified data sets and plotting scripts required to
reproduce the results of this study are openly available at
\url{https://github.com/orlandoue/relativistic-stellar-EPs}.
Within the repository the scripts are in \texttt{code/}, the archived data sets in \texttt{data/} and the figure sources in \texttt{figures/}. The repository contains: the symbolic verification of the cubic discriminant, of the rank-one
Hessian and its kernel at the cusp, and of the migration law \eqref{eq:deborah_migration}
(\path{symbolic_discriminant.py}); the closed-form and multiprecision reproduction of
Table~\ref{tab:certified} (\path{certify_table1.py}); the Chebyshev--hyperboloidal discretisation
of the pencil \eqref{eq:pencil} and its P\"oschl--Teller check at $\zeta_0=0$
(\path{hyperboloidal_pencil.py}); the bordered EP2/EP3 solvers and the $\eta$-continuation of
section~\ref{sec:robustness} (\path{bordered_newton_robustness.py},
\path{branches_Ap_B.py}); the branch-identity test of section~\ref{sec:robustness}(i)
(\path{eigenvector_overlap.py}); the numerical audit reported in Appendix~B, including the
condition numbers, Newton contraction factors and digit stability against $N$ and the working
precision (\path{audit_appendixB.py}); the symbolic verification of the local structure---the expansion \eqref{eq:local_expansion}--\eqref{eq:cusp_unfolding_explicit} with its explicit derivative values, the two counterexamples fixing the scope of the degree-counting argument, the compact forms \eqref{eq:EP_param_compact}, the reality of \eqref{eq:open_real_form}, the common factor \eqref{eq:cusp_cubic} and the resonant divisibility (\path{verify_local_structure.py}); the exact-arithmetic verification of the geometric simplicity of the crossings \eqref{eq:diabolic_family}, carried out in three independent ways---symbolic, rational, and at the level of the discretised operator (\path{code/defective_crossings_proof/}); and the figure generators
(\path{make_all_figures.py}, \path{make_fig4_robustness.py},
\path{make_fig5_postmerger.py}). The archived data sets are
\path{table1_certified.json} (Table~\ref{tab:certified}),
\path{robustness_continuation.json} (branch A, the complex escape, branch R2 and the cusp
trajectory), \path{branches_Ap_B.json} (branches A$'$ and B),
\path{defective_crossings.json} (the exact crossings \eqref{eq:diabolic_family}) and
\path{eigenvector_overlap.json}. Multiprecision quantities are stored as decimal strings, since
they cannot be round-tripped through IEEE doubles, and are archived at the working precision of the
run that produced them. The digits \emph{quoted in the text} are rounded to the accuracy the
conditioning of the bordered Jacobian supports, which is fewer than the digits archived; the archive
is deliberately not truncated to that accuracy, so that the rounding can be checked rather than
assumed.
The repository README gives the figure-to-file map and the order in which the scripts are to be run.

\section*{Conflict of interest}
The author declares no competing interests.

\section*{Funding}
This research received no specific grant from any funding agency in the public, commercial or not-for-profit sectors.

\section*{Acknowledgments}

The author thanks the anonymous referee for a careful and constructive reading of the manuscript across several rounds of revision, which substantially improved the derivation presented in section~\ref{sec:sector_theorem}. The author gratefully acknowledges the developers of the open-source scientific software ecosystem: the multiprecision certification of exceptional points was carried out with \texttt{mpmath}, the symbolic discriminant and unfolding computations with \texttt{SymPy}, and the numerical analysis and figures with \texttt{NumPy}, \texttt{SciPy}, and \texttt{Matplotlib}. Work of this kind depends on freely available scientific software, and on the communities that sustain it.

\medskip
\noindent\rule{\textwidth}{0.4pt}
\medskip

\bibliographystyle{unsrt}
\bibliography{references}

\end{document}